\documentclass[11pt,a4paper]{article}

\usepackage{mystyle}

\definecolor{indigo}{rgb}{0.0, 0.25, 0.75}
{\endgroup}  
\newcommand{\cllbar}{\bar{\mathcal{C}}_{LL}}
\newcommand{\E}{\epsilon}

\newcommand{\sij}[2]{s_{{#1} {#2}}}
\newcommand{\nn}{\nonumber \\}
\newcommand{\mubar}{\bar{\mu}}
\newcommand{\mwsq}{m_W^2}
\newcommand{\tbf}[1]{{\bf T}^{#1}}
\newcommand{\za}[2]{{\langle {#1} {#2} \rangle}}
\newcommand{\zb}[2]{{\left[ {#1} {#2} \right]}}
\newcommand{\zab}[3]{{\left.\langle {#1} | {#2} | {#3} \right]}}

\newcommand{\cll}[2]{\mathcal{C}^{#1}_{LL#2}}
\newcommand{\threef}{3_f }
\newcommand{\amp}[2]{\mathcal{A}^{#1}_{#2}}
\newcommand{\camp}[2]{\mathcal{A}^{^\dagger#1}_{#2}}
\newcommand{\aq}{\bar{q}^{\prime}}

\newcommand{\txb}[1]{\textcolor{indigo}{#1}}
\newcommand{\fac}{\left(\frac{g_{\text{ew}}}{\sqrt{2}}\right)}
\newcommand{\angspinor}[1]{|#1\rangle}
\newcommand{\sqspinor}[1]{|#1]}
\newcommand{\cfq}{\fac} 
\newcommand{\AtsqNLP}[2]{	[\mathcal{\tilde{A}}^2]^{#1}_{#2}|_{\text{NLP}}}
\newcommand{\AsqNLP}[2]{	\left[\mathcal{A}^{#1}_{#2}\right]^2_{\text{NLP}}}
\newcommand{\slog}{\log\left( \frac{s_{45}}{\mubar^2}\right )}
\newcommand{\alog}{\log\left( \frac{s_{12}s_{45}}{s_{13}s_{23}}\right )}
\newcommand{\sqamp}[2]{\left|\mathcal{A}^{#1}_{#2}\right|^2}

\newcommand{\wjet}{$W\!+\!jet$~}
\newcommand{\hjet}{$H\!+\!jet$~}

\newcommand{\dsigma}[2]{\left.	s_{12}^2\frac{d^2\sigma^{#1}_{#2}}{ds_{13}\,ds_{23}}\right|_{\text{NLP-LL}}}

\graphicspath{{images/}}
\allowdisplaybreaks
\begin{document}	
	\begin{titlepage}
		\hypersetup{allcolors=indigo}
		\noindent Preprint \hfill \today
		\vspace*{\fill} \begin{flushleft}
			{\Large \bf {NLP threshold corrections to \wjet production} } 
		\end{flushleft}
		\bigskip
		\medskip 
		\hrule height 0.05cm
		\begin{flushleft}
			\vspace{0.5cm}	
			\bfseries\sffamily {Sourav Pal$\,^{a} $, Satyajit Seth$ \,^b $}
		\end{flushleft}
		\vspace{0.5cm} 
		{\it {$\,^a$ PRISMA Cluster of Excellence, Institut für Physik, Staudinger Weg 7, \\
			Johannes Gutenberg-Universität Mainz, D - 55099 Mainz, Germany\\
			$\, ^b$ Theoretical Physics Division, Physical Research Laboratory,
			Navrangpura, Ahmedabad 380009, India}} 
		\vspace{0.5cm}
		
		\noindent {\it E-mail}: 
		\href{mailto:palsoura@uni-mainz.de}{palsoura@uni-mainz.de},
		\href{mailto:seth@prl.res.in}{seth@prl.res.in} \\ 
		\bigskip \\
		\noindent {Abstract}: We perform a detailed computation of the helicity-dependent next-to-leading power leading logarithms in \wjet production, originating from next-to-soft gluon radiation and soft (anti-)quark emissions. These contributions are systematically captured via helicity-sensitive spinor shifts and soft quark operators. The resulting expressions exhibit full agreement with a recently proposed universal structure of NLP corrections for processes involving the production of an arbitrary massive colourless final state in association with a jet.
		\vspace*{\fill}
	\end{titlepage}
	\hrule
	\hypersetup{allcolors=black}
	\tableofcontents
	\vspace{0.4cm}
	\hrule 	
	\hypersetup{allcolors=indigo}
\section{Introduction}
The continued lack of evidence for phenomena beyond the Standard Model, despite the accumulation of extensive and high-precision data from the Large Hadron Collider (LHC), further strengthens the Standard Model’s standing as a highly robust and predictive theoretical framework for characterizing the fundamental constituents of matter and their interactions. The persistent concordance between experimental findings and theoretical predictions highlights the Standard Model’s sustained empirical success, despite the existence of several fundamental questions that remain beyond its explanatory reach. As experimental precision continues to advance, there is an increasing imperative to develop refined and highly sensitive theoretical frameworks capable of matching or surpassing the sensitivity of current and forthcoming experimental techniques, in order to keep pace with the evolving frontiers of precision in particle physics. 
In perturbative QCD, enhancing theoretical predictions involves either calculating cross-sections at higher orders in strong coupling, or performing an all-order resummation of certain large logarithmic terms. For processes at the LHC, the differential cross-section near the threshold typically includes delta functions alongside two distinct types of logarithmic contributions. One subset of these logarithms, together with the delta functions corresponding to leading-power (LP) terms, exhibits a universal behaviour. Various resummation methods~\cite{Parisi:1980xd,Curci:1979am,Sterman:1987aj,Catani:1989ne,Catani:1990rp,Gatheral:1983cz,Frenkel:1984pz,Sterman:1981jc,Korchemsky:1993xv,Korchemsky:1993uz,
Becher:2006nr,Schwartz:2007ib,Bauer:2008dt,Chiu:2009mg} have been developed to systematically treat these contributions. A comprehensive comparison of different resummation techniques for LP logarithms can be found in refs.~\cite{Luisoni:2015xha,Becher:2014oda,Campbell:2017hsr}). In contrast, the remaining class of logarithms corresponds to next-to-leading power (NLP) terms, which are comparatively less well understood and form the central focus of the present study.		

Given the significant numerical impacts of the NLP logarithms~\cite{Kramer:1996iq,Ball:2013bra,Bonvini:2014qga,Anastasiou:2015ema,Anastasiou:2016cez,vanBeekveld:2019cks,vanBeekveld:2021hhv,Ajjath:2021lvg}, substantial progress has been made towards understanding these logarithms using infrared structure  of gauge theory amplitudes. 
In the last decade, several methods have been developed to resum these  logarithms~\cite{Grunberg:2009yi,Soar:2009yh,Moch:2009hr,Moch:2009mu,Laenen:2010uz,Laenen:2008gt,deFlorian:2014vta,Presti:2014lqa,Bonocore:2015esa,Bonocore:2016awd,Bonocore:2020xuj,Gervais:2017yxv,Gervais:2017zky,Gervais:2017zdb,Laenen:2020nrt,DelDuca:2017twk,vanBeekveld:2019prq,Bonocore:2014wua,Bahjat-Abbas:2018hpv,Ebert:2018lzn,Boughezal:2018mvf,Boughezal:2019ggi,Bahjat-Abbas:2019fqa,Ajjath:2020ulr,Ajjath:2020sjk,Ajjath:2020lwb,Ahmed:2020caw,Ahmed:2020nci,Ajjath:2021lvg,Kolodrubetz:2016uim,Moult:2016fqy,Feige:2017zci,Beneke:2017ztn,Beneke:2018rbh,Bhattacharya:2018vph,Beneke:2019kgv,Bodwin:2021epw,Moult:2019mog,Beneke:2019oqx,Liu:2019oav,Liu:2020tzd,Boughezal:2016zws,Moult:2017rpl,Chang:2017atu,Moult:2018jjd,Beneke:2018gvs,Ebert:2018gsn,Beneke:2019mua,Moult:2019uhz,Liu:2020ydl,Liu:2020eqe,Wang:2019mym,Beneke:2020ibj,vanBeekveld:2021mxn,Das:2024pac,Das:2025wbj,Bhattacharya:2025rqk,vanBijleveld:2025ekz,Czakon:2023tld,Agarwal:2025dvo,Agarwal:2025owu,Hou:2025ovb}.
However, none of the developed methods can confirm the factorisation of next-to-soft radiation at all orders in perturbation theory. The universality of NLP logarithms is already established in case of production of  colour singlet particles~\cite{DelDuca:2017twk}, however for coloured particles in the final state there exists no unique resummation formula. 
One approach to recognising the patterns in NLP logarithms and developing a general resummation formula involves analysing multiple processes with coloured particles in the final state. 
In this context, NLP logarithms for prompt photon plus jet production were investigated in~\cite{vanBeekveld:2019prq} using a method based on shifting pairs of momenta in the squared non-radiative Born amplitudes.

In prior work~\cite{Pal:2023vec,Pal:2024xnu}, we developed a unified framework that integrates colour-ordered helicity amplitudes with soft theorems from gauge and gravity theories~\cite{Strominger:2013jfa,Casali:2014xpa,Luo:2014wea}, facilitating the systematic computation of NLP logarithms for generic scattering processes. Within this formalism, we computed the leading NLP logarithmic corrections for Higgs production in association with a jet via the gluon fusion channel. To incorporate soft quark contributions at NLP, we introduced a set of soft quark operators in~\cite{Pal:2024eyr}, enabling the derivation of NLP threshold corrections across all partonic subprocesses relevant to Higgs plus one jet production. Furthermore, by exploiting symmetry properties of colour-ordered amplitudes, we established the universality of these logarithmic corrections between scalar and pseudo-scalar Higgs production in association with a jet, thereby providing a comprehensive treatment of NLP effects in both these jet associated processes. 

In~\cite{Pal:2025ffp}, we recently established the existence of a universal structure underlying the production of any colourless particle in association with a jet. Within that work, we established a precise correspondence between the mass factorisation kernels and the coefficients of NLP leading logarithms. This connection was derived through general principles based on the structural properties of squared NLP amplitudes and their phase-space integration. As a proof of concept, two specific helicity configurations -- one corresponding to a next-to-soft gluon emission and the other to a soft quark emission -- were examined, illustrating how the NLP leading logarithms for Higgs plus jet production can be leveraged to obtain analogous results for $W$+jet production in those particular helicity sectors. The goal of the present study is to compute explicitly the NLP logarithmic corrections for $W$+jet production across all possible helicity configurations. This work thus constitutes a rigorous and comprehensive validation of the formalism proposed in~\cite{Pal:2025ffp}, further substantiating the universality of NLP logarithmic structures in colourless particle production processes while associated with a final state jet.

For completeness, we recall that this process has been previously investigated at NLP from two complementary perspectives. The factorization formula developed in~\cite{vanBeekveld:2019cks,DelDuca:2017twk} was verified for $W$+jet production in~\cite{vanBeekveld:2023gio}, while the process was also explored in the context of $N$-jettiness subtractions in~\cite{Boughezal:2019ggi}. Nevertheless, neither of these works provides a comprehensive treatment. In particular, ref.~\cite{vanBeekveld:2023gio} provides only the squared matrix elements at NLP, without addressing the subsequent phase space integration. On the other hand, ref.~\cite{Boughezal:2019ggi} concentrates exclusively on N-jettiness subtractions related to next-to-soft gluon emissions and neglects soft quark contributions, which are known to be relevant at NLP. In both cases, the complexity of the NLP squared matrix elements renders the direct integration over phase space exceedingly complex. In this present work, we circumvent this obstacle by formulating the calculation in terms of helicity amplitudes, thereby enabling a more tractable and efficient evaluation of the phase space integrals.

This paper is structured as follows. In section~\ref{sec:gennlp}, we briefly review the structure of amplitudes describing next-to-soft gluon and soft quark emissions. Section~\ref{sec:wjetgen} focuses on the form of helicity amplitudes relevant to $W$ plus one jet production. 
The corresponding NLP leading logarithmic contributions are derived in section~\ref{sec:NLP}, where we also compare the resulting coefficients to those obtained in the $H$+jet process. Finally, in section~\ref{sec:sum}, we summarise our main findings and outline possible directions for future investigation.

\section{Soft and next-to-soft emissions}
\label{sec:gennlp}
The origin of threshold logarithms at LP can be traced back to soft gluon radiation. In contrast, NLP logarithms receive contributions from two distinct sources: $(i)$ next-to-soft gluon emissions and $(ii)$ soft quark emissions. In what follows, we briefly review the methodologies developed in refs.~\cite{Pal:2023vec,Pal:2024eyr} for computing these two types of contributions systematically.

To account for the effects of next-to-soft gluon radiation, we employ the soft theorem in gauge theory~\cite{Strominger:2013jfa,Luo:2014wea,Casali:2014xpa}, which establishes a relation between the $(n+1)$-particle helicity amplitude and the corresponding $n$-particle non-radiative amplitude at both LP and NLP. Specifically, while considering the emission of a soft gluon with positive helicity, we utilize the holomorphic soft limit. In the holomorphic soft limit, the spinor variables associated with the soft gluon $s$ satisfy $\angspinor s \to \lambda \angspinor s$, $\sqspinor s\to$ fixed, with $\lambda\to0$, and the amplitude factorizes as, 
\begin{align}
	\mathcal{A}_{\,n+1}^{\text{LP+NLP}}\bigg(\left\{ \lambda\angspinor s,\sqspinor s\right\} ,\left\{ \angspinor 1,\sqspinor 1\right\} ,\ldots,\left\{ \angspinor n,\sqspinor n\right\} \bigg) & \,=\,\frac{1}{\lambda^{2}}\frac{\braket{1n}}{\braket{1s}\braket{ns}}\times\mathcal{A}_{\,n}\bigg(\left\{ \angspinor 1,\sqspinor{1^{\prime}}\right\} ,\ldots,\left\{ \angspinor n,\sqspinor{n^{\prime}}\right\} \bigg)\,, \label{eq:LPplusNLP}
\end{align}
where the right-hand side involves the non-radiative $n$-particle amplitude, and the square spinors $\sqspinor{1^\prime}$ and $\sqspinor{n^\prime}$ associated with its external legs are shifted in accordance with the next-to-soft kinematics, as described below,
\begin{equation}
	\sqspinor{1^{\prime}}\,=\,\sqspinor 1+\Delta_{s}^{(1,n)}\sqspinor s\,,\qquad
	\sqspinor{n^{\prime}}\,=\,\sqspinor n+\Delta_{s}^{(n,1)}\sqspinor s\,,\qquad
\Delta_{s}^{(i,j)}=\lambda\frac{\braket{js}}{\braket{ji}}\,.
\label{eq:gen-shifts}	
\end{equation}
It is important to note that, in the next-to-soft limit, the shifts of the anti-holomorphic spinors associated with each colour dipole of the non-radiative amplitude generate an independent and distinct colour ordered amplitude that contributes to the $(n+1)$-particle scattering process. In the case where the radiated gluon carries negative helicity, the square spinors are replaced by their corresponding angle spinors and vice versa. 

An additional source of NLP logarithms arises from the emission of soft quarks or anti-quarks. To compute the corresponding contributions within the framework of colour ordered helicity amplitudes, two soft quark operators were introduced in ref.~\cite{Pal:2024eyr}. These operators are constructed by combining a soft quark with a neighbouring hard coloured particle, effectively forming another hard particle that participates in the non-radiative amplitude. Consequently, the colour ordered $(n+1)$-particle amplitude in the presence of soft quark radiation takes the following form, 
\begin{align}
    \amp{s^{i} h^{j}}{n+1}\,=\,\mathcal{Q}^{s^{i} h^{j}\to c^{k}}\,\amp{c^{k}}{n}\,.\qquad
    \label{eq:sqfeynman}
\end{align}
In this expression, $\mathcal{Q}$ denotes the soft quark operator, $s^{i}$ represents the soft (anti-)quark with helicity $i$ and $h^{j}$ denotes the hard particle with helicity $j$. The resulting clubbed colour particle $c^{k}$ with helicity $k$, is formed by merging the soft and hard particles in the $(n+1)$-particle amplitude. Since the soft quark’s contribution is fully encoded in the operator structure, the helicity and momentum of the composite particle are identified with those of the original hard particle. Accordingly, in the above expression, one has $j=k$. There are two distinct ways to combine a soft quark with a hard particle, leading to the definition of two soft quark operators as follows, 
\begin{align}
	\mathcal{Q}^{q_{s}^{+}\bar{q}_{h}^{-}\to g_{c}^{-}}\,=  \,  -\frac{1}{\braket{sh}}\,, \qquad  
\mathcal{Q}^{g_{h}^{+} q_{s}^{+}\to q_{c}^{+}}\,=  \, -\frac{1}{\braket{sh}}\,.
\label{eq:sqop}
\end{align}
All other relevant non-zero combinations can be derived from these two fundamental operators.

\section{Amplitudes for $W$+1 jet production}
\label{sec:wjetgen}
The general form of the scattering amplitude involving a $W^\pm$ boson, a quark--anti-quark pair, and $n$ gluons can be written as, 
\begin{align}
	\amp{\aq q}{Wn}(\{p_i,h_i,c_i\})\,=\,-i\fac g_s^{n}\sum_{\sigma\in S_{n}}(\tbf{c_{4}}\ldots\tbf{c_{n+3}})_{j_{1}j_2}{}\amp{h_1\,h_2\,h_3\,h_4\,\ldots\,h_{n+3}}{1_{\aq}\,2_{q}\,3_W\,4_g\,\ldots\,(n+3)_g}\,.
	\label{eq:genqaqgW}
\end{align}
Here $S_{n}$ represents $n!$ possible non-cyclic permutations of $n$ gluons in the presence of a $q\bar{q}$ pair, and the SU(3) colour generators are normalised as $ \text{Tr} (\tbf{c_i}\tbf{c_j})=\frac{1}{2} \delta^{c_i c_j}$. The strong coupling is denoted with $g_s$ and the electroweak coupling $g_{\text{ew}}=\frac{e}{\sin{\theta_w}}$, where $e$ is the electromagnetic coupling, and $\theta_w$ is the weak mixing angle. Note that the electric charge of the $W$ boson is implicitly determined by the choice of quark ($q$) and antiquark ($\aq$) charges. Accordingly, its explicit charge is left unspecified throughout this study. The computation of NLP threshold corrections to $W$+1-jet production at hadron colliders begins with the evaluation of the scattering amplitude for a $W$ boson produced in association with four partons, one of which becomes soft. At this order, two distinct partonic configurations contribute: $(i)$ a single quark--anti-quark pair accompanied by two gluons, and $(ii)$ two quark--anti-quark pairs produced in association with the $W$ boson.

In the former case ($\aq q W g g \to 0$), corresponding to $n=2$ in eq.~\eqref{eq:genqaqgW}, the amplitude takes the following form,
\begin{align}
	\amp{\aq q}{W2}(\{p_i,h_i,c_i\})\,=\,-i\fac g_{s}^{2}\left[\left(\tbf{c_{4}}\tbf{c_{5}}\right)_{j_{1}j_{2}}\amp{h_1\,h_2\,h_3\,h_4\,h_5}{1_{\aq}\,2_{q}\,3_W\,4_g\,5_{g}}+\left(\tbf{c_{5}}\tbf{c_{4}}\right)_{j_{1}j_{2}}\amp{h_1\,h_2\,h_3\,h_4\,h_5}{1_{\aq}\,2_{q}\,3_W\,5_g\,4_g}\right]\,,
\end{align}
and square of it can be expressed as, 
\begin{align}
	\sum_{\text{colours}}	\left|	\amp{\aq q}{W2}(\{p_i,h_i,c_i\})\right|^2\,&=\,\frac{1}{4}\fac^{2}g_{s}^{4}\,(N^{2}-1)\bigg\{N\left(\left|\amp{h_1\,h_2\,h_3\,h_4\,h_5}{1_{\aq}\,2_{q}\,3_W\,\,4_g\,5_{g}}\right|^{2}+\left|\amp{h_1\,h_2\,h_3\,h_4\,h_5}{1_{\aq}\,2_{q}\,3_W\,\,5_g\,4_{g}}\right|^{2}\right) \nn
	&-\frac{1}{N}\left|\amp{h_1\,h_2\,h_3\,h_4\,h_5}{1_{\aq}\,2_{q}\,3_W\,\,4_g\,5_g}+\amp{h_1\,h_2\,h_3\,h_4\,h_5}{1_{\aq}\,2_{q}\,3_W\,5_g\,4_g}\right|^{2}\bigg\}\,.
	\label{eq:qaqWggsq}
\end{align} 

For the latter configuration $\aq q W \bar{Q} Q \to 0$, the corresponding colour ordered amplitude is given by, 
\begin{align}
	\amp{\aq q}{W \bar{Q} Q}(\{p_i,h_i,c_i\})\,=\, -i \cfq \, g_s^2\, (\tbf{c_1})_{j_1 j_2} (\tbf{c_1})_{j_3 j_4}\,  \amp{h_1\,h_2\,h_3\,h_4\,h_5}{1_{\aq}\,2_{q}3_W\,4_{\bar{Q}}\,5_{Q}}\,,
	\label{eq:W4q}
\end{align}
and the squared amplitude can be written as,  
\begin{align}
	\sum_{\text{colours}} |\amp{\aq q}{W \bar{Q} Q}(\{p_i,h_i,c_i\})|^2\,=\, \frac{1}{4}\cfq^2 g_s^4\, (N^2-1) \left|\amp{h_1\,h_2\,h_3\,h_4\,h_5}{1_{\aq}\,2_{q}\,3_W\,4_{\bar{Q}}\,5_Q} \right|^2\,,
\label{eq:W4qsq1}
\end{align} 
provided that $Q$ is distinct from both $q$ and $q^\prime$. The helicities of the quarks are fixed by those of the anti-quarks. In the scenario where $Q\in\{q,q^\prime\}$, the squared amplitude takes the modified form as shown below,
\begin{align}
	\sum_{\text{colours}} &|\amp{\aq q}{W \bar{Q} Q}(\{p_i,h_i,c_i\})|^2\,=\, \frac{1}{4}\,\cfq^2 g_s^4\, (N^2-1)  \bigg\{ \left|\amp{h_1\,h_2\,\,h_3\,h_4\,h_5}{1_{\aq}\,2_{q}\,3_W\,4_{\bar{Q}}\,5_{Q}} \right|^2+\left|\amp{h_1\,h_2\,h_3\,h_4\,h_5}{1_{\aq}\,2_{q}\,3_W\,4_{\bar{q_x}}\,5_{q_x}}\right|^2 \nn
	&+\frac{\delta_{h_2h_5}}{N}  \bigg(\amp{h_1\,h_2\,h_3\,h_4\,h_5}{1_{\aq}\,2_q\,3_W\,4_{\bar{Q}}\,5_{{Q}}}\amp{\dagger\,h_1\,h_2\,h_3\,h_4\,h_5}{1_q\,2_{\aq}\,3_W\,4_{\bar{q_x}}\,5_{{q_x}}}+\amp{h_1\,h_2\,h_3\,h_4\,h_5}{1_{\aq}\,2_{q}\,3_W\,4_{\bar{q_x}}\,5_{{q_x}}}\amp{\dagger\,h_1\,h_2\,h_3\,h_4\,h_5}{1_{\aq}\,2_{q}\,3_W\,4_{\bar{Q}}\,5_{Q}} \bigg) 
	\bigg\}\,,
	\label{eq:W4qsq2}	
\end{align} 
where $\amp{h_1\,h_2\,h_3\,h_4\,h_5}{1_{\aq}\,2_{q}\,3_W\,4_{\bar{q_x}}\,5_{q_x}}$ corresponds to $\amp{h_1\,h_2\,h_3\,h_4\,h_5}{1_{\aq}\,5_Q\,3_W\,4_{\bar{Q}}\,2_q}$ when $Q=q$, and 
to $\amp{h_1\,h_2\,h_3\,h_4\,h_5}{4_{\bar{Q}}\,2_q\,3_W\,1_{\aq}\,5_Q}$ when $Q=q^\prime$. 
The Kronecker delta appearing in the second line of the above expression enforces that the sub-leading colour contribution is non-vanishing only when the quark helicities are identical.

To compute the colour ordered helicity amplitudes appearing in eq.~\eqref{eq:qaqWggsq}, one may employ methods that generalise the massless scattering amplitude formalism to the case of massive particles with arbitrary spin, as developed in~\cite{Arkani-Hamed:2017jhn}. Further relations among amplitudes involving massive quarks have been explored in~\cite{Ochirov:2018uyq}. A standard method for handling helicity amplitudes with massive $W$ boson as an external state is the light-cone decomposition, where the massive momentum ($p_3$) is written as a sum of two light-like momenta $p^f_{3}$ and $q$, as given below~\cite{Kleiss:1985yh}, 
\begin{align}
	p_3\,=\,p^f_{3}+\frac{\mwsq}{2\,p_3\cdot q} q\,,
	\label{eq:p3fq}
\end{align} 
where $ p_3^2=\mwsq\,,(p^{f}_{3})^2=0\,, $ and $ q^2=0 $.
This parametrisation facilitates the systematic construction of the polarisation vectors for the massive $W$ boson as follows~\cite{Kosower:2004yz,Schwinn:2007ee}, 
\begin{align}
\mathcal{E}^+_\mu (p_3)\,&=\,\frac{\left\langle q|\gamma_\mu| p^{f}_{3} \right]}{\sqrt{2}\,\braket{q\,p^{f}_{3}}}\,, \, 
\mathcal{E}^-_\mu (p_3)\,=\,\frac{\left[ q|\gamma_\mu| p^{f}_{3} \right\rangle}{\sqrt{2}\,[p^{f}_{3}\,q]}, \,
\mathcal{E}^0_\mu (p_3)\,= \,-\frac{p^{f}_{3}}{m_W}+\frac{m_W}{2p^{f}_{3}\cdot q} q\,\,.
\end{align}
With this setup, the independent colour ordered helicity amplitudes read as, 
\begin{eqnarray}
	\amp{+-+++}{1_{\aq}2_q3_W 4_g 5_g} &=& 
  m_w^2 \frac{\za{2}{q}^2}{\za{1}{5} \za{2}{4} \za{4}{5} \za{q}{3_f}^2 } \,, \nonumber \\
%%%%
  \amp{+-++-}{1_{\aq}2_q3_W4_g5_g} &=& 
 -\frac{\za{2}{5} \zb{1}{4}}{\za{2}{4} \zb{1}{5} \za{q}{3_f} s_{45}} 
 \Bigg\{
 \frac{\zb{1}{3_f} \zab{2}{5}{1} \zab{q}{2+5}{4}}{\zb{1}{4}\, s_{245}}
 +\frac{\za{2}{q} \zab{2}{4}{1} \zab{5}{1+4}{3_f}}{\za{2}{5}\, s_{145}}
 +\za{2}{q} \zb{1}{3_f}
 \Bigg\} \,, \nn 
%%%%
  \amp{+--++}{1_{\aq}2_q3_W4_g5_g} &=& 
 -\frac{\za{2}{3_f}^2}{\za{1}{5} \za{2}{4} \za{4}{5} } \,, \nonumber \\ 
%%%%
 \amp{+--+-}{1_{\aq}2_q 3_W 4_g 5_g}&=& 
 \frac{\za{2}{5} \zb{1}{4}}{\za{2}{4} \zb{1}{5} \za{q}{3_f} s_{45}} 
 \Bigg\{
 \frac{\zb{1}{q} \zab{2}{5}{1} \zab{3_f}{2+5}{4}}{\zb{1}{4}\, s_{245}}
 +\frac{\za{2}{3_f} \zab{2}{4}{1} \zab{5}{1+4}{q}}{\za{2}{5}\, s_{145}}
 +\za{2}{3_f} \zb{1}{q}
 \Bigg\} \,, \nonumber \\
 \amp{+-0++} {1_{\aq}2_q 3_W 4_g 5_g}  &=& 
 - \sqrt{2}\, m_w \frac{\za{2}{q} \za{2}{3_f}}{\za{1}{5} \za{2}{4} \za{4}{5} \za{q}{3_f}} \,,  \nn
%%%%
 \amp{+-0+-} {1_{\aq}2_q 3_W 4_g 5_g}  &=& 
 \sqrt{2}\, m_w \frac{1}{s_{45}} \times \nn &&\Bigg\{
 \frac{\za{2}{5} \zb{1}{4} \zab{2}{q}{1}}{\za{2}{4} \zb{1}{5} s_{q 3_f}}
 - \frac{\za{2}{5}^2 \zb{1}{4}}{\za{2}{4} s_{245}} 
 \Bigg( \frac{\zb{1}{q} \zb{4}{3_f}}{\zb{1}{4} \zb{q}{3_f}} + \frac{s_{1q}}{s_{q 3_f}} \Bigg)
 - \frac{\zb{1}{4}^2 \za{2}{5}}{\zb{1}{5} s_{145}} 
 \Bigg( \frac{\za{2}{q} \za{5}{3_f}}{\za{2}{5} \za{q}{3_f}} + \frac{s_{2q}}{s_{q 3_f}} \Bigg)
 \Bigg\}  \,. \nn  
 \label{eq:nloamp}
\end{eqnarray}

To access the contributions relevant at NLP accuracy, we need to consider the limit in which one of the gluons in the above amplitudes becomes next-to-soft, or alternatively, the quark or anti-quark becomes soft. As discussed in section~\ref{sec:gennlp}, the fundamental ingredients at NLP are the non-radiative helicity amplitudes, and the corresponding set of independent non-radiative amplitudes are given by,
\begin{align}
\amp{+-++}{1_{\aq}2_q3_W4_g} \,&= \, 
- m_w^2 \frac{\za{2}{q}^2}{\za{1}{4} \za{2}{4} \za{q}{\threef}^2 } \,, \nn
\amp{+--+}{1_{\aq}2_q3_W4_g}\,&=\,\frac{\za{2}{3_f}^2}{\za{1}{4} \za{2}{4}} \,, \nn
\amp{+-0+}{1_{\aq}2_q3_W4_g}\,&=\,m_w \frac{\za{2}{q} \za{2}{3_f}}{\za{1}{4} \za{2}{4} \za{q}{3_f}}\,.
\label{eq:loampmmd}
\end{align}
These non-radiative amplitudes contribute not only to the calculation of the colour ordered helicity amplitudes appearing in the squared amplitude expression of eq.~\eqref{eq:qaqWggsq}, but also to those entering the squared amplitude formulations presented in eq.~\eqref{eq:W4qsq1} or  eq.~\eqref{eq:W4qsq2}. It is important to note that for the $gg$ initiated channel, neither the quark nor the anti-quark can be taken soft, since the $ggWg\to0$ process does not occur at leading order.

\section{NLP threshold logarithms}
\label{sec:NLP}
To extract the NLP threshold logarithms, we perform an integration of the colour summed squared amplitudes over the unresolved parton phase space. The three-particle phase space is factorised into two two-particle phase spaces -- one describing the kinematics of the two coloured partons with momenta $p_4$ and $p_5$, and the other corresponding to the production of the $W$ boson recoiling against the effective momentum $(p_4+p_5)$. We employ a phase space parametrisation in $d=(4-2\epsilon)$ dimensions, as outlined below~\cite{Ravindran:2002dc,Beenakker:1988bq},
\begin{eqnarray}
	p_1&=&(E_1,0,\cdots,0,E_1) \,, \nonumber \\
	p_2&=& (E_2,0,\cdots,0,p_3\sin \psi, p_3\cos \psi - E_1) \,, \nonumber\\
	p_3&=&-(E_3,0,\cdots,0,p_3\sin \psi, p_3\cos \psi ) \,, \nonumber \\
	p_4&=&-\frac{\sqrt{s_{45}}}{2} (1,0,\cdots,0,\sin \theta_1 \sin \theta_2,\sin\theta_1\cos \theta_2,\cos \theta_1) \,, \nonumber\\
	p_5&=&-\frac{\sqrt{s_{45}}}{2} (1,0,\cdots,0,-\sin \theta_1\sin \theta_2,-\sin\theta_1\cos \theta_2,-\cos \theta_1)\,. 
	\label{eq:para1} 
\end{eqnarray}

The helicity dependent differential cross section at NLP accuracy can then be expressed as follows~\cite{Pal:2023vec,Pal:2024eyr},
\begin{align}
	\left.	s_{12}^2\frac{d^2\sigma^{h_1 h_2 h_4 h_5}}{ds_{13}ds_{23}}\right|_{\text{NLP}}
	\,&= \, \mathcal{F}_{ab} \left (\frac{s_{45}}{\mubar^2}\right )^{-\epsilon}\,\overline{\mathcal{A}_{\text{NLP}}^2} \,,
	\label{eq:crossx}	
\end{align}
with $\mathcal{F}_{ab}$ taking the following form,
\begin{align}
	\mathcal{F}_{ab}\,=\, \frac{1}{16} \kappa_{ab}\, (N^2-1)\,\fac^2\left( \frac{\alpha_{s}(\mubar^2)}{4\pi}\right )^2\,
	\left(\frac{s_{13}\,s_{23}-m_H^2\,s_{45}}{\mubar^2\,s_{12}}\right )^{-\E} \,. 
\end{align}
Here, $s_{45}$ denotes the threshold variable, and the scale $\mubar^2$ is defined as $\mubar^2= 4\pi e^ {-\gamma_{E}\epsilon} \mu_r^2$. The factor $\kappa_{ab}$ accounts for the colour averaging over the initial states, with $\kappa_{q\bar{q}}=1/N^{2}$ and $\kappa_{qg}=\kappa_{\bar{q}g}=1/(N(N^2-1))$. The angular integrated squared amplitude is defined as follows,
\begin{align}
	\overline{\mathcal{A}_\text{NLP}^2}\,=&\,\int_{0}^{\pi} d\theta_1\, (\sin\theta_1)^{1-2\E}\int_{0}^{\pi}d\theta_2\, (\sin\theta_2)^{-2\E} \AtsqNLP{} \,,
\end{align}
where $\AtsqNLP{}{}$ corresponds to the colour summed squared NLP amplitude for a given helicity configuration. For next-to-soft gluon emission, the NLP colour ordered squared amplitude is defined as  $\AsqNLP{}{} = 2 \,\text{Re} (\amp{}{\rm NLP}\amp{\dagger}{\rm LP})$, whereas for soft quark or anti-quark emission it is given by $\AsqNLP{}{} = (\amp{}{\rm NLP}\amp{\dagger}{\rm NLP})$. Note that, from this point onward, all results presented in this work are understood to implicitly include the sum over the polarisation states of the $W$ boson. This summation allows for the reconstruction of the massive $W$ 
boson momentum ($p_3$) from the two light-like momenta $p_3^f$ and $q$ as outlined in eq.~\eqref{eq:p3fq}, thereby permitting the use of phase space parametrisation introduced in eq.~\eqref{eq:para1}. Consequently, the $W$ boson helicity will not be displayed explicitly in the expressions that follow, and due to the momentum conservation, any dependence on $p_3$ can be omitted as well. Within this framework, the general expression for the scattering cross section after the phase space integration is given by, 
\begin{align}
	& \dsigma{h_1 h_2 h_4 h_5}{} \, = \, \left\{\mathcal{C}_{-1}\frac{1}{\epsilon}  
	+\cll{}{} \slog
	\right\}
	(\amp{h_1h_2h_3h_4}{})^2 \,,
	\label{eq:gensigmaNLP}	
\end{align}
where $\cll{}{}$ represents the coefficient of the leading logarithm at NLP in the threshold expansion.

In our recent work~\cite{Pal:2025ffp}, it was demonstrated that $\mathcal{C}_{-1}=-\cll{}{}$, and that the single-pole $\epsilon$-divergence is cancelled upon the consistent application of mass factorisation. Both the underlying identity and the cancellation have been explicitly verified for each helicity configuration examined herein. In the subsequent subsections, we present the results corresponding to the $\cll{}{}$ contribution for various partonic channels relevant to \wjet production. We further present numerical results for these coefficients, normalized by the corresponding Born-level squared amplitudes (thereby yielding $\cllbar$) and evaluated at representative phase space points. For each partonic channel and helicity configuration -- consistent with the analysis in~\cite{Pal:2025ffp} and investigated in detail in this work -- we recover the non-vanishing leading and sub-leading colour contributions previously obtained for the \hjet process, upon substituting the $W$ boson mass with that of the Higgs boson.

\subsection{$ q\aq $ initiated process}
Two distinct partonic channels contribute to the $q\aq$ initiated processes involving the production of a $W$ boson in association with a hard parton and an additional next-to-soft gluon or soft (anti-)quark. The first channel corresponds to 
\begin{eqnarray}
\aq(p_1)+q(p_2)+W(p_3)+g(p_4)+g(p_5) \rightarrow 0 \,,
\label{eq:qqWgg}
\end{eqnarray}
where one of the gluons is next-to-soft, while the second involves 
\begin{eqnarray}
\aq(p_1)+q(p_2)+W(p_3)+Q(p_4)+\bar{Q}(p_5) \rightarrow 0 \,,
\label{eq:qqWQQ}
\end{eqnarray}
with the soft emission of either the quark $Q$ or the anti-quark $\bar{Q}$. Hereafter, soft or next-to-soft particles are indicated by appending the suffix `$(s)$' to the corresponding particle label.

For the first channel as specified in eq.~\eqref{eq:qqWgg}, when the gluon with momentum $p_5$ is next-to-soft and has positive helicity, the relevant colour ordered squared amplitudes contributing to the colour summed squared amplitude as defined in eq.~\eqref{eq:qaqWggsq}, are given by, 
\begin{align}
	\AsqNLP{+-++}{1_{\aq}2_q4_g5_{g(s)}}\,&=\,\frac{2\, \sij{1}{4} \sij{2}{5}}{\left(\sij{1}{2}+\sij{2}{4}\right) \sij{1}{5} \sij{4}{5}} 
	\sqamp{+-+}{1_{\aq}2_q4_g} \,, \nn 
	\AsqNLP{+-++}{1_{\aq}2_q4_g5_{g(s)}}\,&=\,\frac{2\, \sij{2}{4}}{\left(\sij{1}{2}+\sij{2}{4}\right) \sij{4}{5}}\sqamp{+-+}{1_{\aq}2_q4_g} \,, \nn
	\left[\amp{+-++}{1_{\aq}2_q4_g5_{g(s)}} + \amp{+-++}{1_{\aq}2_q4_g 5_{g(s)}} \right]_{\text{NLP}}^2\,&=\, \frac{2 \sij{1}{2}}{\left(\sij{1}{2}+\sij{2}{4}\right)\sij{1}{5}} 
	\sqamp{+-+}{1_{\aq}2_q4_g}\,.
	\label{eq:sgnlpplus}
\end{align}
Using eq.~\eqref{eq:crossx}, we obtain the NLP leading logarithmic contribution to the cross section corresponding to this specific helicity configuration as,
\begin{align}
	\dsigma{+-++}{1_{\aq}\,2_q \,4_ g\,5_{g(s)}}	
	\,=\, \mathcal{F}_{q\bar{q}}\,&\bigg \{  4 \pi \bigg(N-\frac{1}{N}\bigg) \bigg(\frac{   \sij{1}{2}}{(\sij{1}{2} +\sij{1}{3} ) \sij{2}{3}} \bigg) \slog \bigg\} \left|\amp{+-+}{1_{\aq}\,2_q\,4_g}\right|^2. 
	\label{eq:qaqggWp}
\end{align} 

For the same process, when the gluon helicity is negative, the corresponding colour ordered squared amplitudes take the following form,
\begin{align}
	\AsqNLP{+-+-}{1_{\aq}2_q4_g5_{g(s)}}\,=& \, 
	\frac{1}{\sij{1}{5} \sij{2}{4} (\sij{1}{2}+\sij{2}{4}) \sij{4}{5}} \Bigg(
	\sij{4}{5} \sij{1}{2}^2-(\sij{1}{4} \sij{2}{5}+\sij{2}{4} (3 \sij{1}{5}+\sij{4}{5})) \sij{1}{2} \nn
	&+\sij{2}{4} (\sij{1}{4} \sij{2}{5}-\sij{2}{4} (3 \sij{1}{5}+2 \sij{4}{5}))
	\Bigg)
	\sqamp{+-+}{1_{\aq}2_q4_g} \,, \nn 
	\AsqNLP{+-+-}{1_{\aq}2_q5_{g(s)4_g}}\,=& \,
	\frac{1}{\sij{1}{4} (\sij{1}{2}+\sij{2}{4}) \sij{2}{5} \sij{4}{5}} \Bigg(\sij{4}{5} \sij{1}{2}^2+(\sij{2}{4} (\sij{4}{5}-\sij{1}{5})-3 \sij{1}{4} \sij{2}{5}) \sij{1}{2}\nn
	&-\sij{1}{5} \sij{2}{4}^2 -2 \sij{1}{4}^2 \sij{2}{5}+\sij{1}{4} \sij{2}{4} (2 \sij{1}{5}-\sij{2}{5}+2
	\sij{4}{5})\Bigg)
	\sqamp{+-+}{1_{\aq}2_q4_g} \,, \nn
	\left[\amp{+-+-}{1_{\aq}2_q4_g5_{g(s)}}+\amp{+-+-}{1_{\aq}2_q5_{g(s)4_g}}\right]_{\text{NLP}}^2&\,=\, 
	\frac{(\sij{1}{4}+\sij{2}{4}) }{\sij{1}{4} \sij{1}{5} \sij{2}{4} (\sij{1}{2}+\sij{2}{4}) \sij{2}{5}} \Bigg( 
	\sij{2}{4} (\sij{1}{5} \sij{2}{4}-\sij{1}{4} \sij{2}{5}) - \sij{4}{5} \sij{1}{2}^2 \nn
	&+\frac{\sij{1}{2}\sij{1}{4}\sij{2}{4}}{(\sij{1}{4}+\sij{2}{4}) }\left(\frac{\sij{1}{4}}{\sij{2}{4}} \sij{2}{5} +(3 \sij{1}{5}+\sij{2}{5}+\sij{4}{5})+\frac{\sij{2}{4}}{\sij{1}{4}} (\sij{1}{5}-\sij{4}{5})\right)
	\Bigg)
	\sqamp{+-+}{1_{\aq}2_q4_g} \,.
	\label{eq:sgnlpminus}
\end{align}
The associated NLP leading logarithmic contribution to the cross section for this helicity configuration is then given by,
\begin{align}
	\dsigma{+-+-}{1_{\aq}\,2_q \,4_ g\,5_{g(s)}} &\,=\, \mathcal{F}_{q\bar{q}}\, \bigg \{4\pi \left(N-\frac{1}{N}\right)  \left(\frac{ \sij{1}{3}}{(\sij{1}{2}+\sij{1}{3}) \sij{2}{3}} - \frac{1}{\sij{1}{3}}\right) \slog \nn
	&-\frac{4\pi}{N}  \bigg[ 
	 \left( \frac{1}{\sij{1}{3}} + \frac{1}{\sij{2}{3}} - \frac{2}{\sij{1}{2}+\sij{1}{3}} \right)\alog \bigg]
	\bigg\}	\sqamp{+-+}{1_{\aq}2_q4_g} \,.
	\label{eq:qaqggWm}
\end{align} 

\begin{figure}[tbh]
	\begin{center}	
		\includegraphics[scale=0.3]{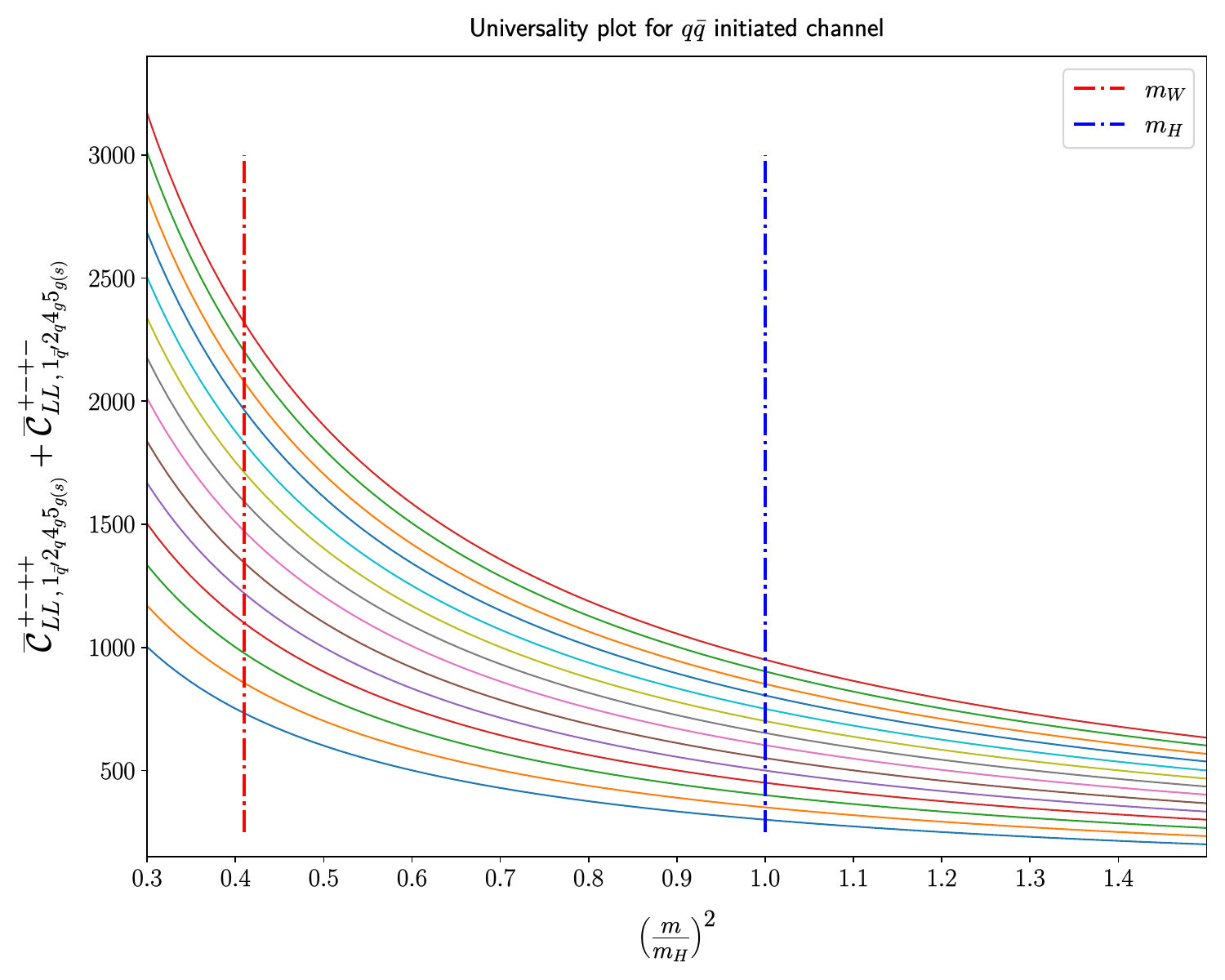} 
		\caption{Plot of $ \cllbar $ for $\aq q$ initiated process with next-to-soft gluon radiation, evaluated at various phase space points. The vertical red line indicates $\cllbar$ obtained by summing the contributions from eqs.~\eqref{eq:qaqggWp} and \eqref{eq:qaqggWm}, while the blue line represents the corresponding result for $ H$ plus one jet production.} 
		\label{fig:qaqWgg}
	\end{center}
\end{figure}

As highlighted in ref.~\cite{Pal:2025ffp}, summing the leading logarithmic coefficients from eq.~\eqref{eq:qaqggWp} and eq.~\eqref{eq:qaqggWm}, 
and factoring out the Born-level squared amplitude, the resulting normalized leading logarithmic coefficient $\cllbar$ precisely reproduces the known result for the \hjet process 
in the identical situation~\cite{Pal:2024xnu}. The only distinction is that, in the \hjet case, the entire contribution arises from the differential distribution of $\sigma^{+-+-}_{1_{\bar{q}}\,2_q \,4_ g\,5_{g(s)}}$ configuration, since the MHV amplitudes does not 
contribute to the NLP corrections due to the next-to-soft gluon emission~\cite{Pal:2024xnu}. In Fig.~\ref{fig:qaqWgg}, we present $\cllbar$ evaluated at several representative phase-space points 
as a function of the colourless heavy particle mass, scaled to the Higgs mass. The continuity between the \hjet and \wjet results under variation of the mass parameter clearly demonstrates 
the universality of the NLP logarithmic structure, as discussed in our previous work~\cite{Pal:2025ffp}.
The NLP leading logarithmic contributions for the remaining helicity configurations can be obtained by exchanging appropriate helicities and/or by permuting the relevant momenta.

In the second channel as defined in eq.~\eqref{eq:qqWQQ}, with $Q=q$, the components necessary for evaluating the colour summed squared amplitude in eq.~\eqref{eq:W4qsq2} are,
\begin{align}
	&\AsqNLP{+-+-}{1_{\aq}2_q4_{\bar{Q}(s)}5_{Q}}\,=\,\frac{\sij{1}{2}}{\sij{1}{5}\sij{4}{5}}\left|\amp{+--}{1_{\aq} 5_q 2_g}\right|^2 \,, \nn
	&\AsqNLP{+-+-}{1_{\aq}2_q4_{\bar{q}(s)}5_{q}}\,=\,\frac{1}{\sij{2}{4}}\left|\amp{+--}{1_{\aq} 5_q 2_g}\right|^2 \,, \nn
	&\left[\amp{+-+-}{1_{\aq}2_q4_{\bar{Q}(s)}5_{Q}} \camp{+-+-}{1_{\aq}2_q4_{\bar{q}(s)}5_{q}} + c.c. \right]_{\text{NLP}} = 
	\left( \frac{\sij{1}{4} \sij{2}{5}}{\sij{1}{5} \sij{2}{4} \sij{4}{5}} - \frac{\sij{1}{2}}{\sij{1}{5} \sij{2}{4}} -\frac{1}{\sij{4}{5}} \right) \left|\amp{+--}{1_{\aq} 5_q 2_g}\right|^2 \,,
	\label{eq:amp4qq}
\end{align} 
and upon applying eq.~\eqref{eq:crossx}, one can express the NLP leading logarithmic contribution to the cross section as,
\begin{align} 
	\dsigma{+-+-}{1_{\aq}2_q4_{\bar{q}(s)}5_{q}}\,
	&=\,\mathcal{F}_{\aq q}\bigg \{4\pi N\left(\frac{1}{2\sij{1}{3}}\right)\slog
	\bigg\} \left|\amp{+--}{1_{\aq} 5_q 2_g}\right|^2 \,.
	\label{eq:xsec4qq}
\end{align}

In the case where $Q=q^\prime$, the analogous expressions in eqs.~\eqref{eq:amp4qq} and \eqref{eq:xsec4qq} are modified accordingly and take the following form,
\begin{align}
	&\AsqNLP{+-+-}{1_{\aq}2_q4_{\bar{Q^\prime}}5_{Q^\prime(s)}}\,=\,\frac{\sij{1}{2}}{\sij{2}{4}\sij{4}{5}}\left|\amp{+-+}{4_{\aq} 2_q 1_g}\right|^2 \,,\nn
	&\AsqNLP{+-+-}{1_{\aq}2_q4_{\aq}5_{q^\prime(s)}}\,=\,\frac{1}{\sij{1}{5}}\left|\amp{+-+}{4_{\aq} 2_q 1_g}\right|^2 \,, \nn
	&\left[\amp{+-+-}{1_{\aq}2_q4_{\bar{Q^\prime}}5_{Q^\prime(s)}} \camp{+-+-}{1_{\aq}2_q4_{\aq}5_{q^\prime(s)}} + c.c. \right]_{\text{NLP}} = 
	\left( \frac{\sij{1}{4} \sij{2}{5}}{\sij{1}{5} \sij{2}{4} \sij{4}{5}} - \frac{\sij{1}{2}}{\sij{1}{5} \sij{2}{4}} -\frac{1}{\sij{4}{5}} \right) \left|\amp{+-+}{4_{\aq} 2_q 1_g}\right|^2 \,,
	\label{eq:same-flav}
\end{align} 
and
\begin{align} 
	\dsigma{+-+-}{1_{\aq}2_q4_{\aq}5_{q^\prime(s)}}\,
	&=\,\mathcal{F}_{\aq q}\bigg \{4\pi N\left(\frac{1}{2\sij{2}{3}}\right)\slog
	\bigg\} \left|\amp{+-+}{4_{\aq} 2_q 1_g}\right|^2 \,.
	\label{eq:diff-flav}
\end{align}

\begin{figure}[htb]
	\centering
	\subfloat[][]{\includegraphics[scale=0.3]{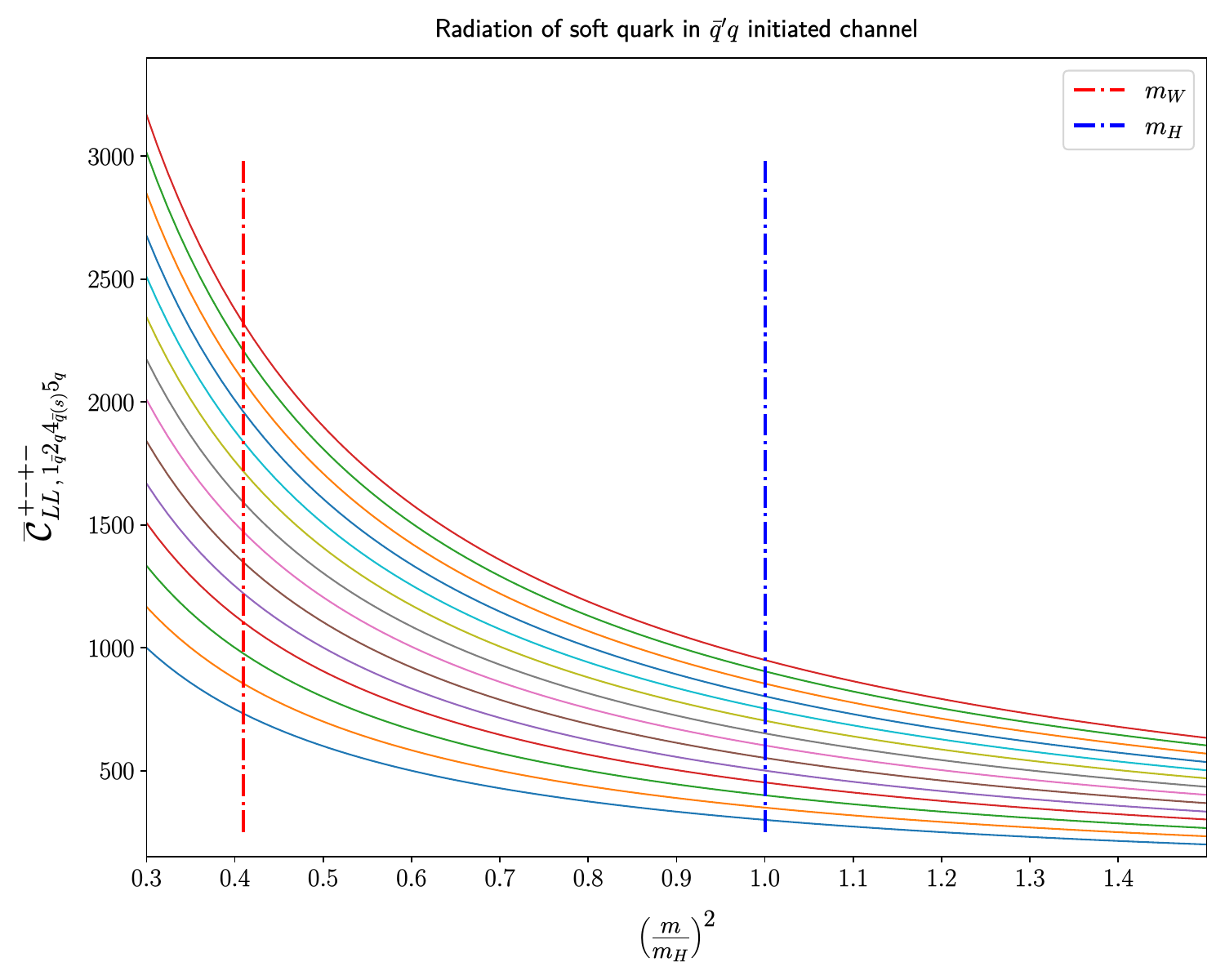} }
	\subfloat[][]{\includegraphics[scale=0.3]{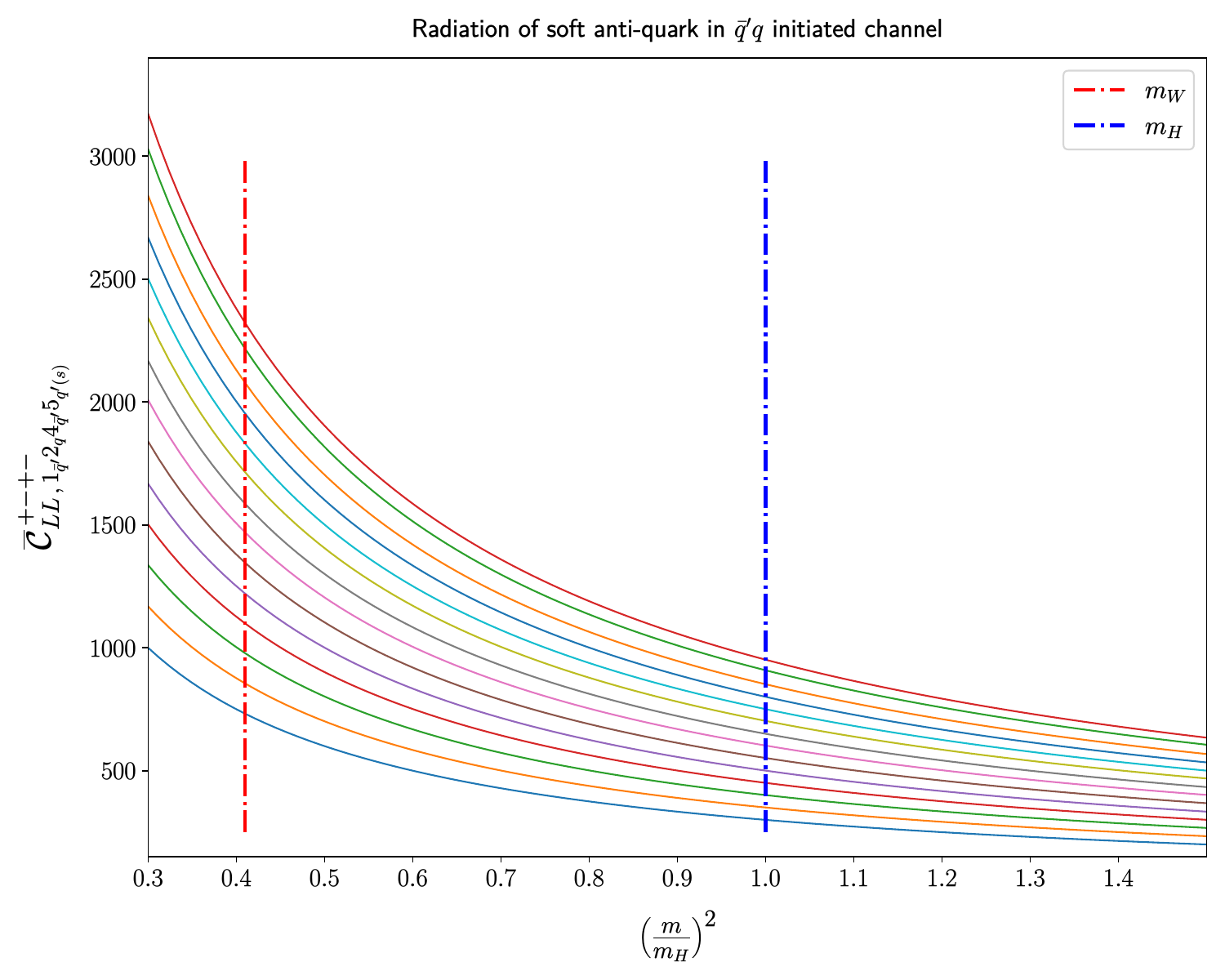} } \\
	\caption{Panel~(\txb{a}) shows the $\cllbar$ result for radiation of a final state soft quark for $Q=q$, as given in eq.~\eqref{eq:xsec4qq}. Panel~(\txb{a}) displays the corresponding result for radiation of a final state soft anti-quark in the case $Q=q^\prime$, as given in eq.~\eqref{eq:diff-flav}. In both panels, the vertical red lines indicate the results for the \wjet process, while the vertical blue lines correspond to the results for \hjet production.}
	\label{fig:aqq-cll}
\end{figure}

We observe that no NLP leading-logarithmic contributions arise when $Q\notin\{q,q^\prime\}$, since the non-vanishing term exclusively stems from the second component of the leading colour term, as shown in the second line of eq.~\eqref{eq:W4qsq2}, but absent in eq.~\eqref{eq:W4qsq1}. Moreover, the sub-leading colour term involving the Kronecker delta, presented in the second line of eq.~\eqref{eq:W4qsq2}, does not yield any NLP leading logarithmic contribution. 
This further accounts for the omission of NLP leading logarithmic contributions associated with the differential cross sections of $\sigma^{+-+-}_{1_{\aq}2_q4_{\bar{q}}5_{q}(s)}$ and $\sigma^{+-+-}_{1_{\aq}2_q4_{\aq(s)}5_{q^\prime}}$ in the above, as these contributions vanish, owing to the fact that the soft particle clubs exclusively to the immediately adjacent anti-particle in only one permitted manner, thereby effectively reproducing the scenario while $Q\notin\{q,q^\prime\}$.
Fig.~\eqref{fig:aqq-cll} displays various non-zero $\cllbar$ coefficients, illustrating the anticipated universal behaviour of the Born squared amplitude normalized NLP leading logarithmic contributions, as noted in ref.~\cite{Pal:2025ffp}. 

\subsection{$gg$ initiated process}
The only contributing process to the $gg$ initiated channel is, 
\begin{align}
	g (p_1) +g (p_2) +W^-(p_3) + \aq (p_4) + q (p_5) \rightarrow 0 \,.
	\label{eq:ggWqq}	
\end{align}  
For the NLP leading logarithmic contributions relevant to the \wjet process, either the final state quark or the anti-quark may become soft. We begin by examining the case where the quark with momentum $p_5$ is soft. In this configuration, two independent helicity amplitudes emerge, with the associated colour ordered squared amplitudes entering eq.~\eqref{eq:qaqWggsq} explicitly detailed as follows,
\begin{align}
	\AsqNLP{+++-}{1_g2_g4_{\aq}5_{q(s)}}\,&=0 \,, \nn
	\AsqNLP{+++-}{1_g2_g4_{\aq}5_{q(s)}}\,&=0 \,, \nn  
	\left|\amp{+++-}{1_g2_g4_{\aq}5_{q(s)}}+\amp{+++-}{2_g1_g4_{\aq}5_{q(s)}}\right|_{\text{NLP}}^2\,&=0 \,,
\end{align}
and
\begin{align}
	\AsqNLP{+-+-}{1_g2_g4_{\aq}5_{q(s)}}\,&=\,0 \,, \nn
	\AsqNLP{-++-}{2_g1_g4_{\aq}5_{q(s)}}\,&=\, \frac{1}{\sij{2}{5}} \sqamp{+-+}{4_{\aq}2_q1_g} \,, \nn
	\left|\amp{+-+-}{1_g2_g4_{\aq}5_{q(s)}}+\amp{-++-}{2_g1_g4_{\aq}5_{q(s)}}\right|_{\text{NLP}}^2\,&=\,\frac{1}{\sij{2}{5}} \sqamp{+-+}{4_{\aq}2_q1_g} \,.
	\label{eq:ggsq}
\end{align}
As previously indicated, the polarisations of the $W$ boson are summed over and consequently suppressed throughout this work. Following the phase space integration over the unresolved particle in accordance with eq.~\eqref{eq:crossx}, the NLP leading logarithmic contributions for these two independent helicity configurations can be expressed as,
\begin{align}
	\dsigma{+++-}{1_g2_g4_{\aq}5_{q(s)}}\,&=\, 0 \,, \nn
	\dsigma{+-+-}{1_g2_g 4_{\aq}5_{q(s)}}\,&=\, \mathcal{F}_{gg}\, \bigg \{4\pi \left(N-\frac{1}{N}\right) \left(\frac{1}{2\sij{1}{3}}
	\right) \slog \bigg\} \sqamp{+-+}{4_{\aq} 2_q 1_g} \,.
	\label{eq:ggcllsq}
\end{align}

In the alternative scenario where the anti-quark with momentum $p_4$ becomes soft, the colour ordered squared amplitudes and their associated NLP leading logarithmic contributions are summarized as follows, 
\begin{align}
	\AsqNLP{+++-}{1_g2_g4_{\aq(s)}5_{q}}\,&=\,\frac{1}{\sij{2}{4}} \, \sqamp{+-+}{2_{\aq}5_q1_g} \,, \nn  
	\AsqNLP{+++-}{1_g2_g4_{\aq(s)}5_{q}}\,&=\, \frac{\sij{1}{5}}{\sij{1}{4} \sij{2}{5}} \sqamp{+-+}{2_{\aq}5_q1_g} \,, \nn 
	\left[\amp{+++-}{1_g2_g4_{\aq(s)}5_{q}}+\amp{+++-}{1_g2_g4_{\aq(s)}5_{q}}\right]\,&=\, \frac{\sij{1}{2} \sij{4}{5}}{\sij{1}{4} \sij{2}{4} \sij{2}{5}} \sqamp{+-+}{2_{\aq}5_q1_g} \,, \nn 
	\label{eq:ggWsaq1}
\end{align}
\begin{align}
	\dsigma{+++-}{1_g2_g4_{\aq(s)}5_{q}}\,&=\, \mathcal{F}_{gg}\, \bigg \{ 4\pi \bigg[  \left(N-\frac{1}{N}\right) \slog -\frac{1}{N} \alog \bigg] \nn
	&\qquad\times \left(\frac{1}{2 \sij{1}{3}} \sqamp{+-+}{2_{\aq}5_q1_g} + \frac{1}{2 \sij{2}{3}} \sqamp{+-+}{1_{\aq}5_q2_g} \right) 
	  \bigg\} \,, \nn
	\label{eq:ggcllsaq1}
\end{align}
and 
\begin{align}
	\AsqNLP{+-+-}{1_g2_g4_{\aq(s)}5_{q}}\,&=\,0 \,, \nn
	\AsqNLP{-++-}{2_g1_g4_{\aq(s)}5_{q}}\,&=\, \frac{1}{\sij{1}{4}} \sqamp{+--}{1_{\aq}5_q2_g} \,, \nn
	\left|\amp{+-+-}{1_g2_g4_{\aq(s)}5_{q}}+\amp{-++-}{2_g1_g4_{\aq(s)}5_{q}}\right|_{\text{NLP}}^2\,&=\,\frac{1}{\sij{1}{4}}\sqamp{+--}{1_{\aq}5_q2_g} \,,
	\label{eq:ggWsaq2}
\end{align}
\begin{align}
	\dsigma{+-+-}{1_g2_g4_{\aq(s)}5_{q}}\,&=\, \mathcal{F}_{gg}\, \bigg \{4\pi \left(N-\frac{1}{N}\right) 
	\left(\frac{1}{2 \sij{2}{3}}
	\right) \slog  \bigg\} \sqamp{+--}{1_{\aq}5_q2_g} \,.
	\label{eq:ggcllsaq2}
\end{align}

\begin{figure}[htb]
	\centering
	\subfloat[][]{\includegraphics[scale=0.3]{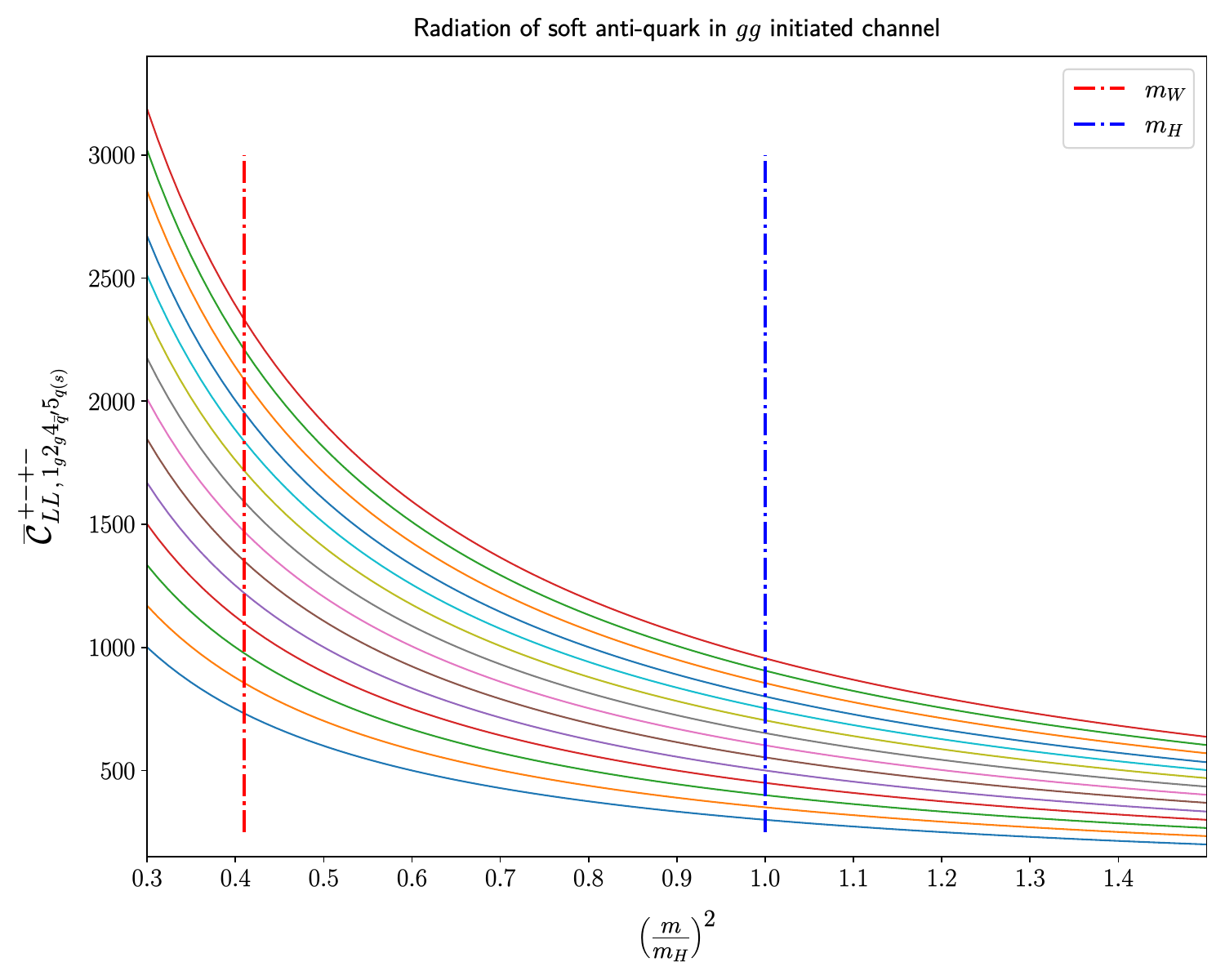} }
	\subfloat[][]{\includegraphics[scale=0.3]{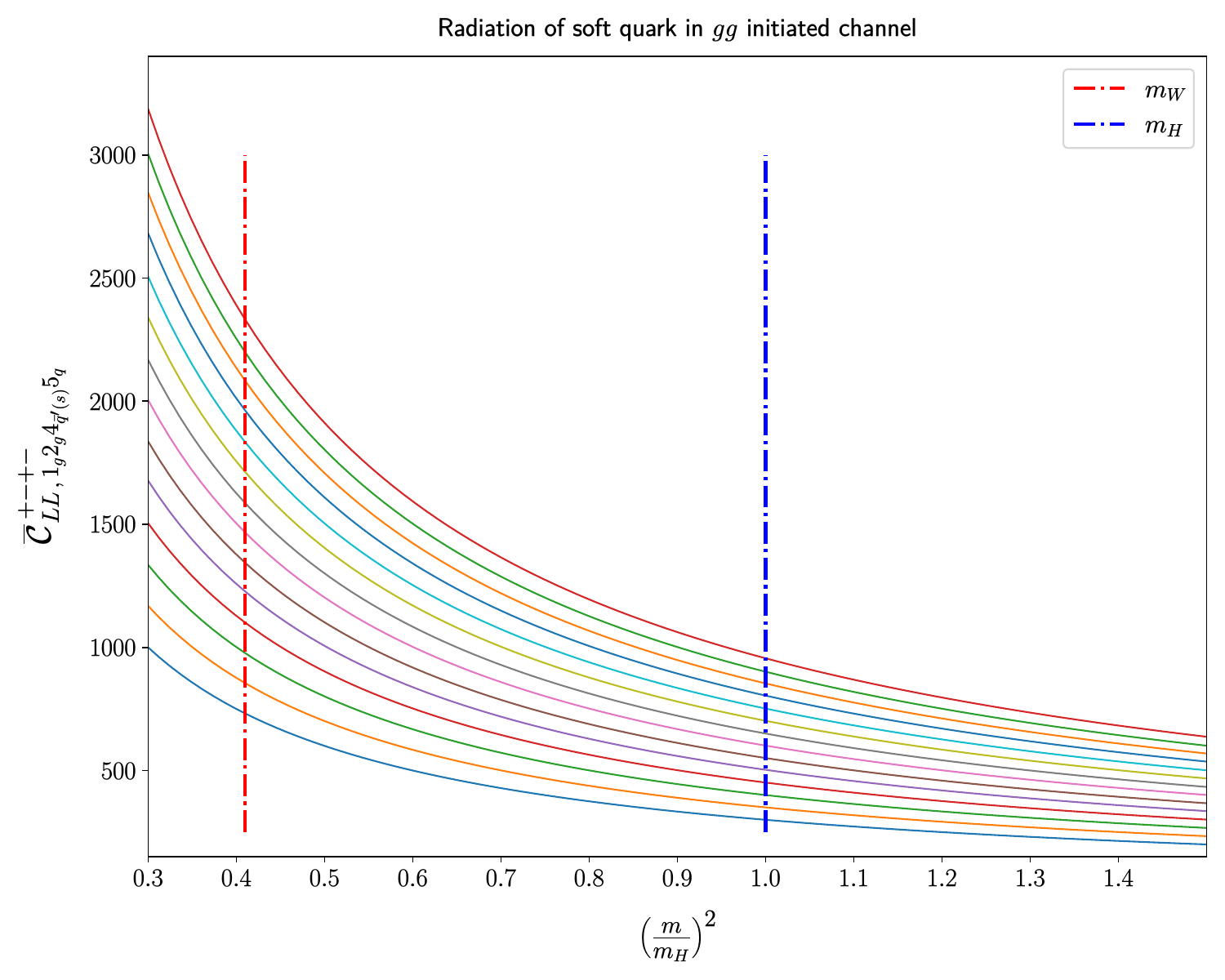} }
	\caption{The $\cllbar$ coefficients arising from final state soft quark and anti-quark radiation are shown in panels (a) and (b), corresponding to eqs.~\eqref{eq:ggcllsq} and \eqref{eq:ggcllsaq2} respectively. In both plots, the vertical red lines represent the results for the \wjet process, while the vertical blue lines denote the corresponding results for \hjet production. The result associated with eq.~\eqref{eq:ggcllsaq1} is not displayed explicitly, as its behaviour can be readily inferred from the results presented in these two panels.}
	\label{fig:gg-cll}
\end{figure}

The normalised non-vanishing NLP leading-logarithmic coefficients $\cllbar$, as obtained from eqs.~\eqref{eq:ggcllsq} and \eqref{eq:ggcllsaq2}, are presented in Fig.~\ref{fig:gg-cll}. 
Eq.~\eqref{eq:ggcllsaq1} contains two Born-level squared amplitudes whose corresponding $\cllbar$ structures are analogous to those obtained from eqs.~\eqref{eq:ggcllsq} and \eqref{eq:ggcllsaq2}. Therefore, this result is not shown explicitly in this figure, as it can be straightforwardly deduced from the $\cllbar$ shown in Fig.~\ref{fig:gg-cll}(\txb{a}) and Fig.~\ref{fig:gg-cll}(\txb{b}).
As in previous cases, we demonstrate how these results extend to the \hjet process by varying the mass of the massive colourless particle, thereby reinforcing the universality argument put forward in ref.~\cite{Pal:2025ffp}.

\subsection{$qg$ initiated process} 
The process contributing to the $qg$ initiated channel proceeds via 
\begin{align} 
	q (p_1) + g(p_2) + W^-(p_3) + g (p_4) +\aq (p_5) \rightarrow 0 \,,
\end{align} 
wherein either of the final state partons can become (next-to-)soft, thereby yielding NLP leading logarithmic contributions. In the configuration where the gluon with momentum $p_4$ becomes next-to-soft, the relevant squared matrix elements can be obtained through a straightforward relabelling of momenta in eqs.~\eqref{eq:sgnlpplus} and \eqref{eq:sgnlpminus} as given below,
\begin{align} 
	\Big\{1\to5, 2\to1, 4\to2, 5\to4 \Big\}\,.  
\end{align} 
Upon integrating over the phase space of the unresolved parton according to eq.~\eqref{eq:crossx}, the NLP leading logarithmic contributions corresponding to the distinct helicity configurations are found to be, 
\begin{align} 
	\dsigma{-+++}{1_q2_g\,4_{g(s)}\,5_{\bar{q}^\prime}}\,=\,  \mathcal{F}_{qg} \bigg\{ & 4 \pi N \bigg [\left (\frac{2\sij{1}{2}}{\sij{1}{3}(\sij{1}{2}+\sij{2}{3})} \right) \slog  \bigg] \bigg \} \left|\amp{+-+}{5_{\bar{q}^\prime} 1_ q2_g}\right|^2 \,, \nn
	\dsigma{-+-+}{1_q2_g\,4_{g(s)}\,5_{\bar{q}^\prime}}\,=\,\mathcal{F}_{qg} \bigg \{ & 4\pi N \bigg [\left
	(\frac{4}{\sij{1}{3}}-\frac{1}{\sij{2}{3}} - \frac{2\sij{1}{2}}{\sij{1}{3}(\sij{1}{2}+\sij{2}{3})}\right) \slog  \nn
	& + \left( \frac{1}{\sij{1}{3}} - \frac{2}{\sij{1}{2}+\sij{2}{3}} \right) \alog \bigg] \nn 
	& + \frac{4\pi}{N} \bigg [\left(\frac{1}{\sij{2}{3}} \right) \bigg] \slog\bigg \} \left|\amp{+-+}{5_{\bar{q}^\prime} 1_ q 2_g}\right|^2 \,.
	\label{eq:qgsg}
\end{align}
The results corresponding to helicity configurations related by helicity reversal remain identical to those presented above. The remaining configurations can be obtained by suitably relabelling the external momenta. 
 
We now turn to the complementary kinematic configuration in which the final state quark with momentum $p_5$ becomes soft. In this case as well, the relevant colour ordered squared amplitudes can be obtained by applying the momentum relabelling 
\begin{align}
	\{5\rightarrow1,1\rightarrow2,2\rightarrow4,4\rightarrow5\}
\end{align} 
to eqs.~\eqref{eq:ggWsaq1} and \eqref{eq:ggWsaq2}. The resulting next-to-leading power leading logarithmic contributions to the differential cross section for the independent helicity configurations can then be expressed as,
\begin{align} 
	\dsigma{-+++}{1_ q2_g\,4_g\,5_{\aq(s)}}\, &=\,\mathcal{F}_{qg} \bigg \{ 4 \pi \bigg(N-\frac{1}{N}\bigg) \bigg [\left (\frac{1}{2\sij{1}{3}}\right) \slog  \bigg] 
        \bigg \} \left|\amp{+-+}{2_{\aq} 1_ q4_g}\right|^2 \,, \nn 
	\dsigma{-+-+}{1_q2_g4_g5_{\aq(s)}}\,&=\,\mathcal{F}_{qg} \bigg \{ 4 \pi \bigg(N-\frac{1}{N}\bigg) \bigg [\left (\frac{1}{2\sij{1}{3}}\right) \slog  \bigg] 
        \bigg \} \left|\amp{+--}{2_{\aq} 1_ q4_g}\right|^2 \,.
	\label{eq:qgsaq}	 
\end{align} 
\begin{figure}[htb]
	\centering
	\subfloat[][]{\includegraphics[scale=0.3]{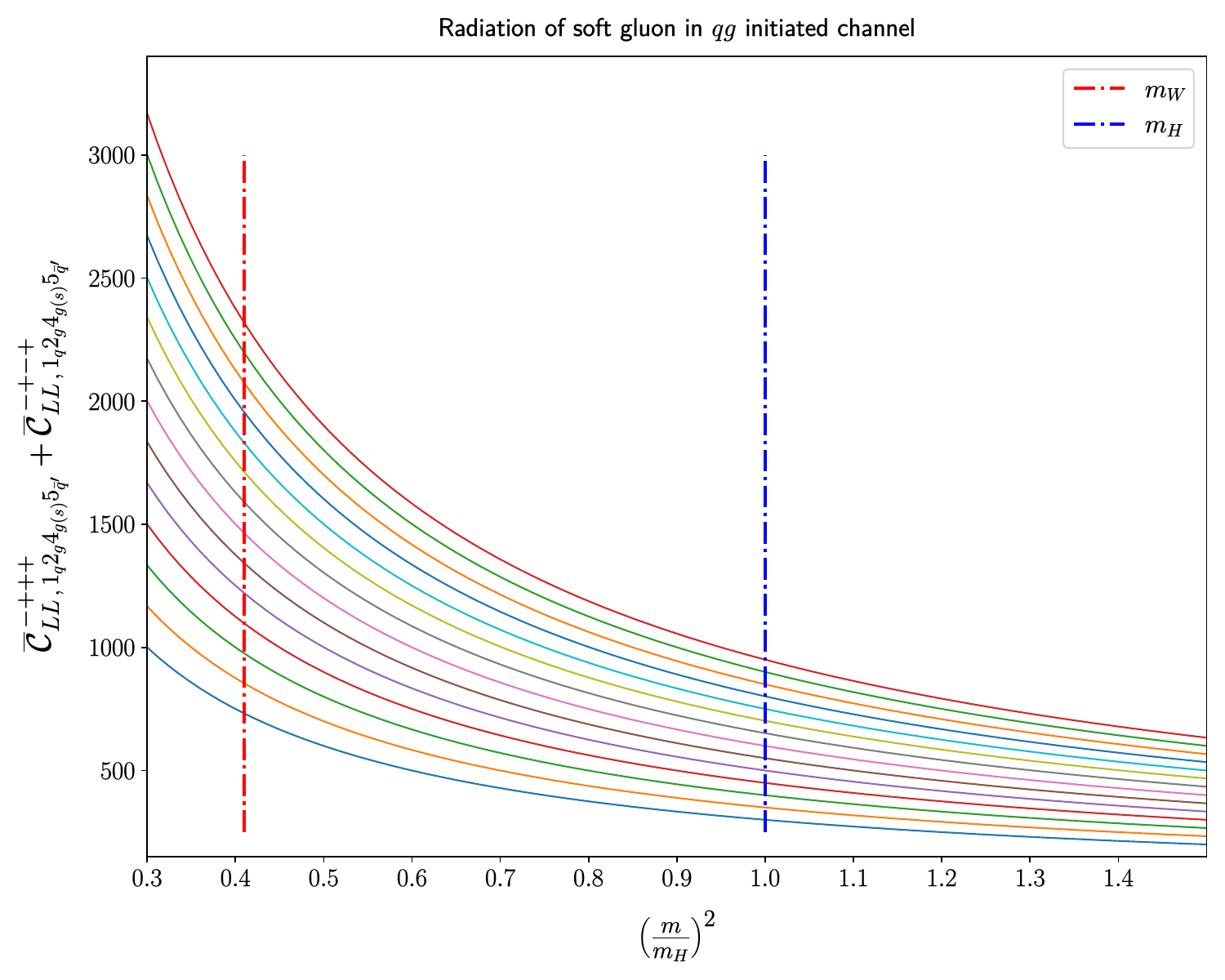} }
	\subfloat[][]{\includegraphics[scale=0.3]{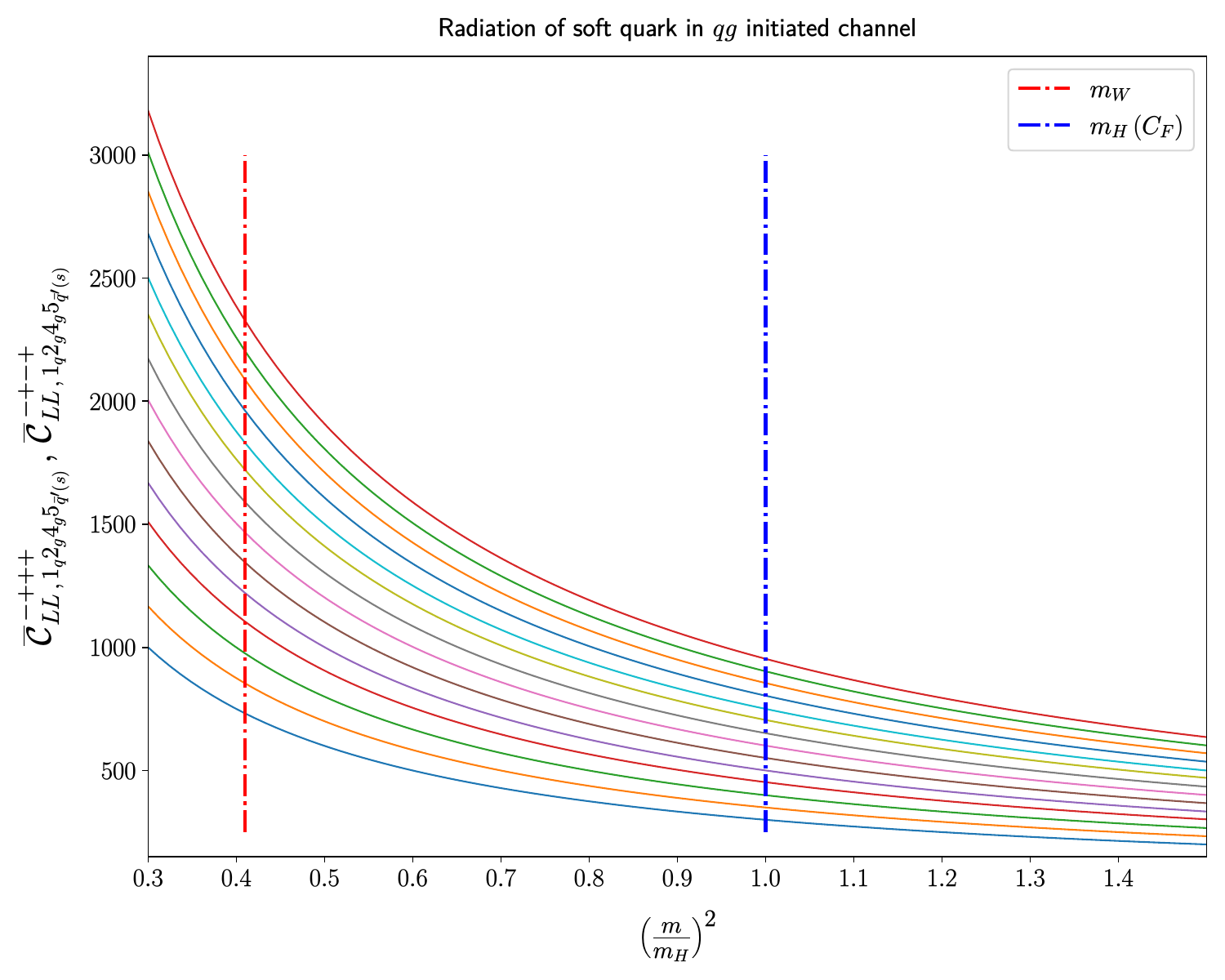} }
	\caption{Panel~(\txb{a}) shows the $\cllbar$ result for the radiation of an unpolarised next-to-soft gluon, obtained by summing both the expressions in eq.\eqref{eq:qgsg}. Panel~(\txb{b}) presents the $\cllbar$ for soft final state quark emission, including two helicity configurations, which yield identical expressions as given in eq.~\eqref{eq:qgsaq}. In both panels, the vertical red lines indicate the results for the \wjet process. The vertical blue line in panel~(\txb{a}) corresponds to the full result for \hjet production. In contrast, the blue line in panel~(\txb{b}) shows only the partial \hjet contribution proportional to $C_F$, as additional contribution arises in this case due to the non-vanishing $ggHg\to0$ non-radiative process.}
	\label{fig:qg-cll}
\end{figure}
Fig.~\ref{fig:qg-cll} depicts the mass dependence of the normalized $\cllbar$ coefficients evaluated at representative phase space points for this $qg$ initiated channel. Panels (\ref{fig:qg-cll}\txb{a}) and (\ref{fig:qg-cll}\txb{b}) display results obtained from the NLP leading logarithmic coefficients associated with soft quark radiation, as defined in eq.~\eqref{eq:qgsaq} within the context of \wjet production. 
When mapped onto the \hjet process, this contribution accounts solely for the partial component proportional to  $C_F=\frac{N^2-1}{2 N}$. This limitation arises because, in contrast to \hjet, the \wjet channel does not feature leading order process such as $ggWg\to0$ that yield contributions proportional to $C_A=N$. Nevertheless, as evidenced by the results in panel (\ref{fig:qg-cll}\txb{c}), the full \hjet prediction can be recovered from the corresponding \wjet calculation in the case of next-to-soft gluon emission. Collectively, these findings provide robust support for the universality scenario advanced in ref. ~\cite{Pal:2025ffp}.

\subsection{$ \aq g $ initiated process} 
For the $\aq g$ initiated channel, the process under consideration is given by, 
\begin{align}
	g(p_1)+ \aq(p_2)+W^-(p_3)+q(p_4)+g(p_5)\rightarrow 0 \,.
	\label{eq:qaq}	
\end{align} 
In analogy with the $qg$ initiated case discussed in the preceding subsection, the pertinent colour ordered squared amplitudes entering eq.~\eqref{eq:qaqWggsq} can be derived through a straightforward relabelling of external momenta. In the case of next-to-soft gluon emission, the appropriate momentum mapping is 
\begin{align}
	\Big\{1\to2, 2\to4, 4\to1 \Big\} \,,
\end{align} 
which needs be applied to the expressions in eqs.~\eqref{eq:sgnlpplus} and \eqref{eq:sgnlpminus}. Upon computing the colour summed squared amplitudes, the NLP logarithms follow directly from eq.~\eqref{eq:crossx}, and are explicitly given by, 
\begin{align} 
	\dsigma{++-+}{1_g2_{\aq}4_q5_{g(s)}}\,&=\,\mathcal{F}_{qg}\bigg \{ 4 \pi N \bigg [\left(\frac{3}{\sij{1}{3}+\sij{2}{3}}\right) \slog  +\left(\frac{2}{\sij{1}{3}+\sij{2}{3}}\right) \alog \bigg] \nn 
	& -\frac{ 4 \pi}{N} \bigg [\left (\frac{1}{\sij{1}{3}+\sij{2}{3}}\right)
	\slog \bigg]  \bigg \} \left|\amp{+-+}{2_{\aq} 4_ q 1_g}\right|^2. \nn
	\dsigma{++--}{1_g2_{\aq}4_q5_{g(s)}}\,&=\,\mathcal{F}_{qg} 
	\bigg \{ 4 \pi N \bigg[\left (
	\frac{4}{\sij{2}{3}}+\frac{1}{\sij{1}{3}}-\frac{3}{\sij{1}{3}+\sij{2}{3}}\right)\slog
	\nn & 
	+\left(\frac{1}{\sij{2}{3}}-\frac{2}{\sij{1}{3}+\sij{2}{3}}\right) \alog  \bigg]
	\nn & 
	- \frac{ 4 \pi}{N} \bigg [\left ( \frac{1}{\sij{1}{3}}-\frac{1}{\sij{1}{3}+\sij{2}{3}}
	\right) \slog \bigg]  \bigg\} \left|\amp{+-+}{2_{\aq} 4_ q 1_g}\right|^2.
	\label{eq:aqgsg}  
\end{align} 

Similarly, for the emission of the final state soft anti-quark, the relabelling 
\begin{align}
	\{1\rightarrow2,2\rightarrow4,4\rightarrow5,5\rightarrow1\}
\end{align}
applied to eqs.~\eqref{eq:ggWsaq1} and \eqref{eq:ggWsaq2} yields the relevant colour ordered squared amplitudes for this configuration. Integration over the unobserved phase space of the soft anti-quark then produces the NLP leading logarithmic contributions corresponding to the independent helicity configurations as presented below, 
\begin{align} 
	\dsigma{++-+}{1_g2_{\aq}4_{q(s)}5_{g}}\,&=\,0 \nn
	\dsigma{-+-+}{1_g2_{\aq}4_{q(s)}5_{g}}\,
	&=\,\mathcal{F}_{qg}\bigg \{4\pi \bigg(N-\frac{1}{N}\bigg) \bigg
	[\left(\frac{1}{2\sij{2}{3}}\right)\slog
        \bigg]\bigg\}\left|\amp{+-+}{2_{\aq} 1_ q 5_g}\right|^2.
	\label{eq:aqgsq}
\end{align}
\begin{figure}[htb]
	\centering
	\subfloat[][]{\includegraphics[scale=0.3]{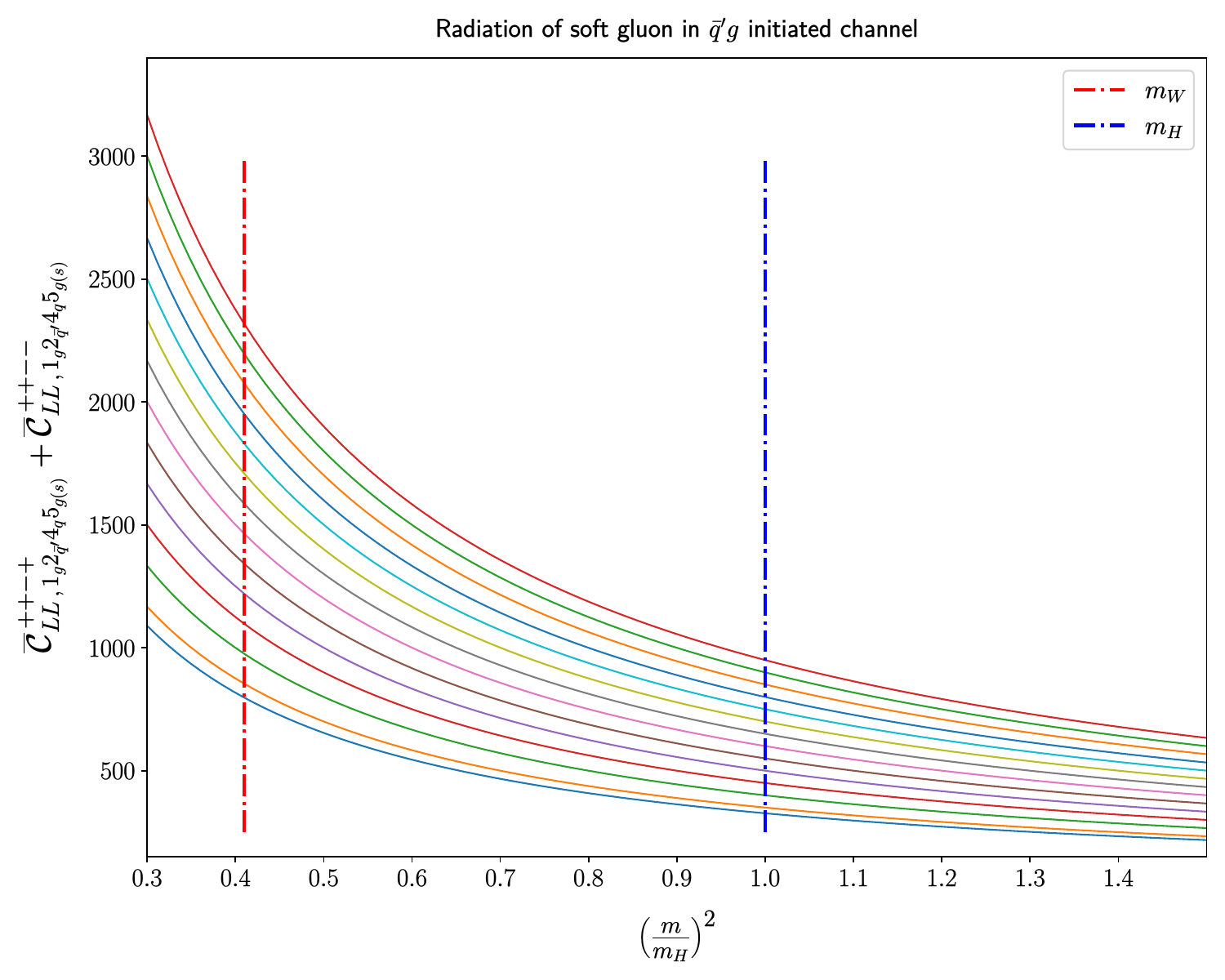} }
	\subfloat[][]{\includegraphics[scale=0.3]{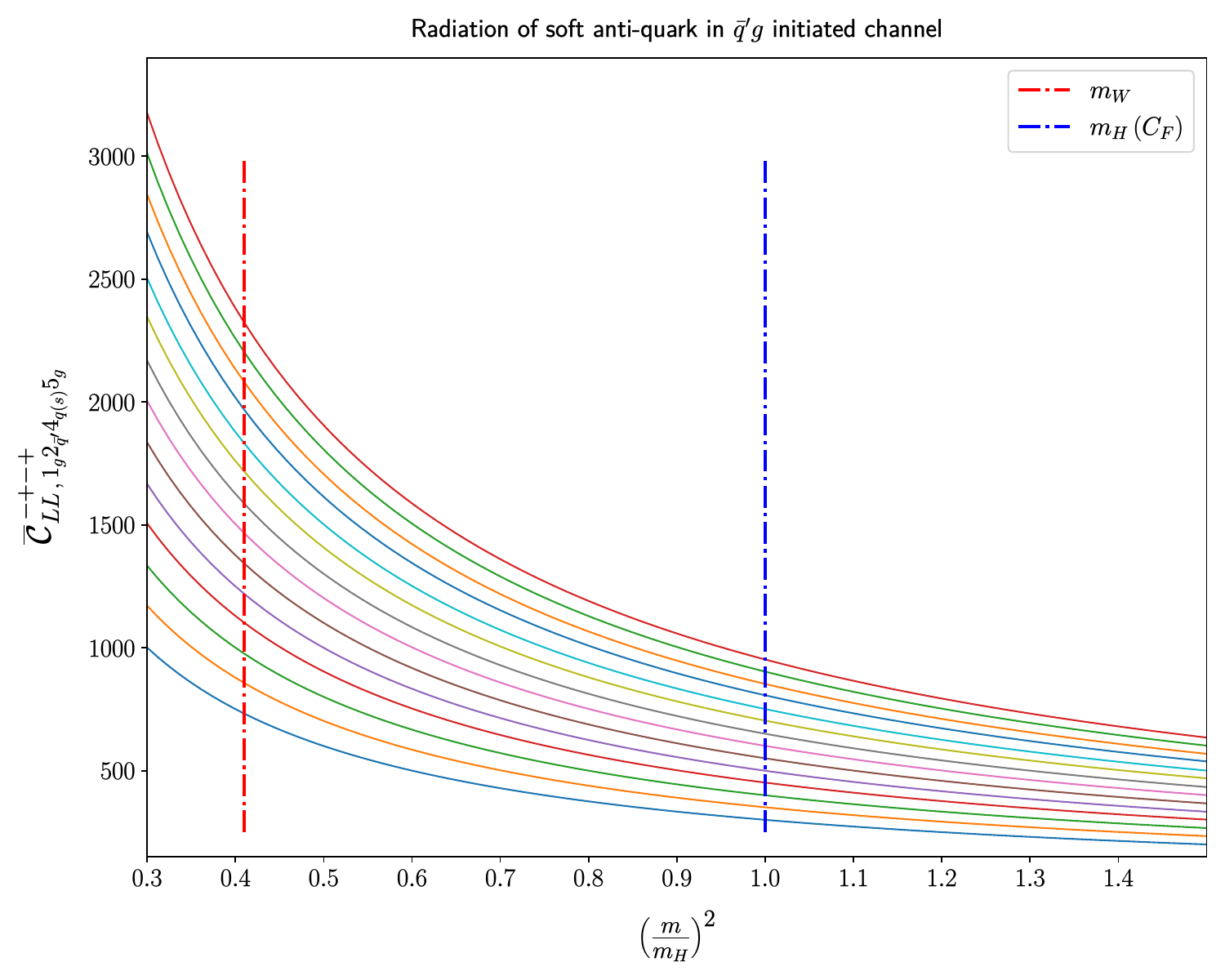} } \\
	\caption{Panel~(\txb{a}) shows the $\cllbar$ result for next-to-soft gluon radiation, with both polarisation states summed, while panel~(\txb{b}) presents the $\cllbar$ for radiation of a soft final state anti-quark. In both panels, the vertical red lines indicate the results for the \wjet process, corresponding to eq.~\eqref{eq:aqgsg} (sum of both contributions) in panel~(\txb{a}), and eq.~\eqref{eq:aqgsq} in panel~(\txb{b}). The vertical blue line in panel~(\txb{a}) denotes the result for \hjet production. In panel~(\txb{b}), the blue line represents the partial \hjet result proportional to $C_F$, as the $C_A$ contribution is absent in the \wjet case due to the absence of non-radiative $ggWg\to0$ process at leading order.}
	\label{fig:aqg-cll}
\end{figure}
Fig.~\ref{fig:aqg-cll} illustrates the variation of the Born square normalized coefficients $\cllbar$ corresponding to next-to-soft gluon and soft quark emissions, evaluated over a range of phase space points and mass values. As previously discussed, unlike the \hjet process, the \wjet production does not involve a $gg$ initiated leading order partonic channel and consequently, the soft anti-quark contribution in \wjet differs from that in \hjet by a term proportional to $C_A$, as noted in ref.~\cite{Pal:2025ffp}. As anticipated, the term proportional to $C_F$ is identical in both cases. For next-to-soft gluon radiation, the normalized NLP leading logarithmic coefficients can be smoothly translated to those of \hjet case, as demonstrated in panel (\ref{fig:aqg-cll}\txb{a}), thereby further substantiating the universality framework initially identified in ref.~\cite{Pal:2025ffp}. 

Before concluding this section, we note in passing that the same universality extends to prompt photon plus jet production, once the photon polarisations are summed over. This can be readily inferred from the helicity-driven differential cross sections presented above -- upon summing over the relevant helicity configurations where appropriate, the resulting expressions of the leading logarithmic coefficients are manifestly independent of $\sij{1}{2}$ and $m_W$, thereby pointing to the existence of a smooth $m_W\to 0$ limit.

Furthermore, to quantify the impact of NLP effects, we study the relative contributions of the NLP and LP leading logarithmic contributions at the partonic level across a large sample of phase-space points for each helicity configuration. We find that including NLP leading logarithmic corrections can modify the partonic LP results by up to $25\%$ in certain kinematic regions. The larger deviations predominantly arise from the emergence of distinct leading and sub-leading colour contributions at NLP, highlighting the potential relevance of these effects for achieving the percent-level precision targeted at the LHC. We emphasize, however, that this estimate is obtained at the partonic level, and the inclusion of parton distribution functions, realistic jet-clustering algorithms, and other experimental effects may alter the result. A detailed phenomenological study incorporating these ingredients is deferred to future work.

\section{Summary}
\label{sec:sum}
The absence of unambiguous signals of physics beyond the Standard Model in the high-precision dataset accumulated at the LHC highlights the pressing need for a more refined theoretical understanding of the Standard Model itself. Achieving the requisite level of theoretical precision for jet associated production processes necessitates advancements in the understanding of NLP threshold corrections, which arise from emissions of next-to-soft gluons or soft (anti-)quarks near the Born kinematics, as these corrections contribute significantly to the cross section. We compute the NLP corrections associated with next-to-soft gluon emissions using the method of shifted spinors as developed in ref.~\cite{Pal:2023vec}. In parallel, the contributions arising from soft (anti-)quark emissions are evaluated via operator-based methods formulated in the spinor-helicity framework, following the formalism introduced in ref.~\cite{Pal:2024xnu}.

The principal aim of this study is to validate the universality principle for NLP leading logarithms proposed in ref.~\cite{Pal:2025ffp}, in the context of the production of a massive, colour-singlet final state in association with a jet. The framework presented therein suggests that the NLP leading logarithmic coefficients admit a universal representation derived via mass factorization, with process independent splitting kernels capturing the relevant dynamics. Importantly, this universality is conjectured to hold independently of the spin of the massive colourless particle, provided that all relevant helicity configurations are correctly accounted for.

Focusing on the case of \wjet production, we confirm the proposed universal formula for all helicity configurations and reproduce the results previously obtained for \hjet production in ref.~\cite{Pal:2024xnu}. Our analysis demonstrates that the NLP leading logarithmic contributions from soft quark emissions maintain a coherent analytic structure, contingent upon the inclusion of all relevant Born level squared amplitudes and a sum over the polarisations of the final state W boson. Analogously, for emissions of next-to-soft gluons, the NLP threshold leading logarithmic terms exhibit a universal analytic structure once summed over the gluon helicities and the polarisations of the W boson. 
Nevertheless, this universality can be extended to massless final states accompanied by a jet, such as prompt photon–jet production, provided that photon polarisations are summed and the appropriate helicity configurations are identified. Moreover, the coefficients of the NLP leading logarithms remain robust under the inclusion of any additional next-to-soft effects arising from phase-space constraints. 

%\begin{report}
%Furthermore, we have verified that the results for $ \gamma$+jet production can be directly recovered from our $W $+ jet expressions by taking the smooth limit $ m_W\rightarrow 0$, which makes the universality of the logarithms more robust. 
%
%
%We also note that an additional source of NLP logarithms arises from next-to-collinear gluon emissions. While the present work focuses exclusively on the soft-emission regime, the universal structure of the amplitudes for arbitrary boson-plus-jet production and the consistent phase-space transition from massive to massless bosons strongly suggests that a similar universality may extend to the next-to-collinear sector. We defer a rigorous investigation of these collinear effects to a dedicated future study.
%
%As we are interested in the processes with final state jets, therefore while applying jet clustering algorithms, we will encounter the clustering and non-global logarithms due to phase-space restrictions.  However the $\cll{}{}$s' presented in this paper remain unaffected as the phase space restrictions start contributing at NLL and beyond at next-to-leading power.
%   
%\end{report}

Our findings thus substantiate the claim of the universal nature of NLP logarithms, and reinforce the utility of helicity-driven NLP analyses for the development of systematic resummation frameworks. In particular, our findings facilitate the incorporation of NLP effects into resummation of observables for processes involving the production of colour-singlet particles in association with a jet, with potential applications to both polarised and unpolarised cross sections. It would be worth investigating whether analogous universal patterns emerge in the case of collinear emissions and beyond the leading logarithms in NLP.

\section*{Acknowledgements}
The research of SP is supported from
the Alexander von Humboldt Foundation. 
The work of SS is supported by the Department of Space, Government of India. SS  acknowledges partial support by the ANRF/SERB MATRICS under Grant No. MTR/2022/000135.

\bibliographystyle{bibstyle}
\bibliography{Wjet.bib}

\end{document}